\documentclass[11pt]{article}   	
\usepackage{geometry}                		
\usepackage{graphicx}				
\usepackage{amssymb}
\usepackage{amsmath}
\usepackage{authblk}
\usepackage[table]{xcolor}
\usepackage{multirow}
\usepackage{graphicx}
\usepackage[normalem]{ulem}
\usepackage[english]{babel}
\usepackage{appendix}
\usepackage{float}
\usepackage{xcolor}
\usepackage{siunitx}
\usepackage{comment}

\newcommand{\be}{\begin{equation}}
\newcommand{\ee}{\end{equation}}

\usepackage[utf8]{inputenc}
\usepackage{xspace}
\usepackage{lineno}

\usepackage[hypertexnames,
    pdftex,%
    colorlinks,%
		citecolor=blue,%
    hyperindex,%
    plainpages=false,%
    bookmarksopen,%
    bookmarksnumbered%
  ]{hyperref}
\usepackage[numbers,sort&compress]{natbib}
\definecolor{eM}{hsb}{0.3,0.5,1}
\definecolor{eS}{hsb}{0,1,0.6}

\graphicspath{{figs/}}

\usepackage{color}
\usepackage[normalem]{ulem}
\definecolor{pncolor}{rgb}{0,0.1,0.7}
\definecolor{ascolor}{rgb}{1,0,1}
\definecolor{nacolor}{rgb}{1,0,0}
\definecolor{wscolor}{rgb}{0,0.6,0.2}

\DeclareRobustCommand\pnout{\bgroup\markoverwith{\color{pncolor}{\rule[0.4ex]{2pt}{0.8pt}}}\ULon}
\DeclareRobustCommand\asout{\bgroup\markoverwith{\color{ascolor}{\rule[0.4ex]{2pt}{0.8pt}}}\ULon}
\DeclareRobustCommand\naout{\bgroup\markoverwith{\color{nacolor}{\rule[0.4ex]{2pt}{0.8pt}}}\ULon}
\DeclareRobustCommand\wsout{\bgroup\markoverwith{\color{wscolor}{\rule[0.4ex]{2pt}{0.8pt}}}\ULon}

\title{\bf 
Inclusive electron-proton measurement prospects in the Electron-Ion Collider early science stage
}
\author[1]{Javier Jim\'enez-L\'opez \thanks{javiej09@ucm.es (Corresponding author)}}
\author[2]{Stephen Maple \thanks{s.maple@bham.ac.uk}}
\author[3]{Paul R. Newman \thanks{p.r.newman@bham.ac.uk}}
\author[4]{Katarzyna Wichmann \thanks{katarzyna.wichmann@desy.de}}
\affil[1]{\small \it Departamento de F\'isica Te\'orica, Universidad Complutense de Madrid, E-28040 Madrid, Spain}
\affil[2,3]{\small \it School of Physics and Astronomy, University of Birmingham, B15 2TT, UK}
\affil[4]{\small \it Deutsches Elektronen-Synchrotron DESY, Germany}
\date{
DESY-25-164 \\
November 2025}
\begin{document}
\maketitle
\begin{abstract}
We explore the potential for extracting proton 
structure functions, proton parton density functions (PDFs), and the strong coupling $\alpha_s(M_z^2)$,
using early science data from the future Electron-Ion Collider (EIC),
both standalone, and in combination with HERA data. 
Different scenarios
are considered in which samples with modest luminosity 
are collected at either two or three EIC beam energy configurations. 
The Rosenbluth separation method 
is used to extract the proton structure functions 
$F_2$ and $F_L$ from simulated data
in a model-independent manner, showing that $F_L$ can
be extracted significantly more precisely with three centre
of mass energies than with two, whilst also obtaining $F_2$  
to higher precision than has been achieved previously. 
The inclusion of a third beam configuration is also beneficial 
in the extraction of the strong coupling $\alpha_s(M_z^2)$ that is
obtainable with unprecedented  experimental precision with the early EIC data.
Additionally, the precision of the proton PDFs is improved when adding these data, especially for large values of Bjorken-$x$, for both two and three EIC beam energy configurations.
These studies show that EIC data will already be a
highly competitive probe of perturbative Quantum Chromodynamics within
the first five years of data taking.
\end{abstract}

\section{Introduction}
\label{sec:intro}

The internal structure of the proton is 
an important topic in high-energy physics, offering 
insights into the composition and dynamics of 
strongly interacting matter. Central to these studies is the measurement of cross sections for Deep Inelastic Scattering (DIS) 
and thus of the proton structure functions, which in turn
provide insight into the distributions of quarks and gluons 
in the nucleon \cite{10.1093/acprof:oso/9780198506713.001.0001}. The Hadron-Electron Ring Accelerator (HERA) was the first and, to date, the only high energy 
electron-proton colliding beam facility. It allowed measurements of inclusive Deep Inelastic Scattering (DIS) cross sections, and extraction of the structure functions $F_{2}$, $F_{L}$ and $x F^{\gamma Z}_3$\cite{H1:2015ubc}. Among these, 
the generalised structure function, $F_2$, is closely related to the quark density in the proton, whilst the gluon density can be obtained from the scaling violations of $F_2$ or from the longitudinal structure function $F_L$~\cite{Alekhin:2017kpj,Hou:2019efy,Bailey:2020ooq,NNPDF:2021njg,Cooper-Sarkar:1987cnv,Boroun:2012bje,Zijlstra:1992qd}.
Knowledge of both the $F_2$ and $F_L$ 
structure functions enables
extractions of the 
photoabsorption ratio $R$, which corresponds to the cross section ratio for 
longitudinally to transversely polarised virtual photons
interacting with the proton, such that

The extractions of $F_{L}$ at HERA \cite{H1:2013ktq,ZEUS:2014thn} are limited by statistics and cover a restricted range at low values of the Bjorken variable $x$.
The $F_L$ structure function has also been measured at low $Q^2$ and large $x$ by numerous fixed-target
DIS experiments \cite{BENVENUTI1989485,ARNEODO19973,E140X:1995ims,Tvaskis:2006tv,Whitlow:1991uw,PhysRevC.97.045204,JeffersonLabHallCE94-110:2004nsn}.
The Electron-Ion Collider~\cite{Accardi:2012qut} (EIC), 
which is expected to begin science operations at Brookhaven 
National Laboratory in the mid 2030s,
promises to revolutionise our understanding of the 
internal structure of hadrons. 
Intended to operate at high luminosities and with a wide range of beam energy configurations, the 
EIC will provide the opportunity to extract the proton
longitudinal structure function with unprecedented precision 
over a wide kinematic range in $x$ and 
$Q^2$~\cite{Jimenez-Lopez:2024hpj}.

The EIC project plan is divided such that the EIC science programme may commence before the full capabilities of the collider are realised. The ``early science'' period, is foreseen for the first five years 
of EIC operation and will provide access to a few selected centre-of-mass energies for electron-proton and electron-ion scattering data, with limited luminosities. 
At present, two different electron-proton beam energy configurations are anticipated for early science. 
In this study, we consider different approaches to extracting the $F_2$ and $F_{L}$ 
structure functions using EIC early science data with two centre-of-mass energies,
including extractions that combine HERA and EIC data.
Additionally, we investigate the
improvements that may be possible with the addition of a third centre-of-mass energy in the first five years.
Further studies are conducted to evaluate the impact of EIC early science inclusive data on determinations of parton distribution functions (PDFs) and the strong coupling $\alpha_s$.

\section{EIC pseudo-data} \label{pseudo}

The accelerator and detector designs for the EIC are currently undergoing intense development, but the overall specification in terms of beam energy ranges and 
instrumentation coverage and performance
are already well-established~\cite{AbdulKhalek:2021gbh}. 
The studies presented in this paper are based on simulated data points or `pseudo-data'
which are derived from that baseline configuration. 
Our approach to producing inclusive DIS EIC pseudo-data 
follows that of \cite{Jimenez-Lopez:2024hpj,Armesto:2023hnw}, which in turn took 
binning schemes based on those in
the ATHENA detector proposal~\cite{ATHENA:2022hxb}. The ATHENA 
collaboration has since merged with ECCE~\cite{adkins2023design} to create the 
ePIC collaboration, which is now working
towards a first EIC detector. 
While some specifics of the detectors have evolved, the overall
expected kinematic range, kinematic variable resolutions and achievable experimental
precision are largely independent of the detailed detector design and this
study is thus applicable to any general purpose EIC experiment. 

The EIC ``early science'' period, corresponding approximately to the first five years of data taking, 
is currently planned to include two distinct EIC $ep$ beam energy settings, achieved via the scattering of \SI{10}{\giga\electronvolt} electrons on \SI{130}{\giga\electronvolt} or \SI{250}{\giga\electronvolt} protons. 
The early science EIC will, however, be capable of operating with \SI{5}{\giga\electronvolt} electron beams as well. When combined with the \SI{130}{\giga\electronvolt} and \SI{250}{\giga\electronvolt} proton beams this offers two additional beam energy configurations, $5\times\SI{130}{\giga\electronvolt\squared}$ and $5\times\SI{250}{\giga\electronvolt\squared}$. The $5\times\SI{250}{\giga\electronvolt\squared}$ and $10\times\SI{130}{\giga\electronvolt\squared}$ configurations are effectively degenerate in centre-of-mass energy. 
However, the $5\times\SI{130}{\giga\electronvolt\squared}$ beam setting offers a technically possible third point, 
which is additionally included in our studies. 
We therefore consider four scenarios: EIC
pseudo-data only with either two or three beam energy settings, and HERA data combined with pseudo-data for either two or three EIC configurations.
As summarised in Table~\ref{tab:samples}, the EIC pseudo-data are thus produced for an
integrated luminosity of 1~fb$^{-1}$, corresponding to conservative
expectations for a few months of early science
data collection, 
at three different centre-of-mass energies $\sqrt{s}$. 
Since we explore the possible extraction of
the structure functions using EIC pseudo-data together with the final HERA inclusive DIS cross sections,
the EIC $Q^2$ and $x$ bin centres are chosen to match those of~\cite{H1:2015ubc} for the $e^+p$ scattering at four different centre-of-mass energies\footnote{These bins do not correspond directly to the dedicated $F_L$ measurements at HERA~\cite{H1:2013ktq,ZEUS:2014thn} but to the general binning chosen for the combination of the H1 and ZEUS inclusive cross sections.}. 
For each beam configuration, the pseudo-data are produced at five logarithmically  (but not evenly) spaced $x$ values per decade over the 
inelasticity range \(0.005 < y < 0.96\), 
matching the expected experimental resolutions~\cite{ATHENA:2022hxb}.
The resolution of ePIC in \(Q^2\) is expected to be significantly better than that of the HERA, so the HERA $Q^2$ binning is adopted. 

\begin{table}[htb!]
\centerline{
\begin{tabular}{|c|c|c|c|}
\hline
$e$-beam energy (GeV) & $p$-beam energy (GeV) & $\sqrt{s}$ (GeV) & Integrated lumi (fb$^{-1}$) \\ \hline 
10 & 250 & 100 &  1\\
10 & 130 & 72 &  1\\
5 &  130 &  51 &  1 \\ \hline
\end{tabular}}
\vskip 0.4cm
\caption{Beam energies, centre-of-mass energies and 
integrated luminosities assumed for the 
different EIC configurations considered for early science.
}
\label{tab:samples}
\end{table}

The central values of the pseudo-data cross sections are 
obtained using 
NNLO theoretical predictions based on the HERAPDF~\(2.0\) parton densities~\cite{H1:2015ubc} within the xFitter framework~\cite{Alekhin:2014irh,H1:2009pze,H1:2009bcq},
with the $Q^2$ evolution performed according to the
NNLO DGLAP evolution equations~\cite{Gribov:1972ri,Gribov:1972rt,Lipatov:1974qm,Dokshitzer:1977sg,Altarelli:1977zs,Moch:2004pa,Vogt:2004mw,Almasy:2011eq,Mitov:2006ic,Blumlein:2021enk,Ablinger:2024qxg,Ablinger:2025awb}. 
The value for each data point is
randomly smeared using samples from Gaussian distributions that reflect the assumed 
experimental uncertainties.

Taking into account that
not all systematic uncertainties are known with certainty in this time several years before the commencement of early science operations,
we consider the ``conservative scenario'' for the uncertainties 
introduced in~\cite{Jimenez-Lopez:2024hpj}.
It follows the considerations in~\cite{AbdulKhalek:2021gbh}, 
which were also adopted in the studies by  
the ATHENA collaboration, and subsequently used to study EIC collinear PDF sensitivities~\cite{Armesto:2023hnw} and \(\alpha_s\) measurements~\cite{Cerci:2023uhu}. 
The data points are attributed a point-to-point uncorrelated systematic uncertainty 
of 1.9\(\%\),
determined by the sum in quadrature of a conservative 1~\% from radiative corrections, and 1.6~\% from other effects such as bin migration, detector efficiency and charge symmetric background. 
A normalisation uncertainty
of 3.4\(\%\) is also included for each data set, which is not correlated between 
different beam energy configurations, such that the total systematic uncertainty 
attributed to each data point is 3.9 \(\%\). 
Here, the normalisation uncertainties comprise a 1.5~\% contribution from the luminosity measurement, with the remaining 3~\% arising from electron purity and detector efficiency effects.

The statistical uncertainties corresponding to 1~fb$^{-1}$ of integrated luminosity were confirmed in dedicated studies to be negligible compared to the systematic uncertainties,
for all but the bins at the very largest $x$ and $Q^2$.
This remains the case after including efficiency
and resolution effects at a level that might be 
expected at the start of EIC operations.


\section{Structure functions $F_2$ and $F_{L}$}
\label{sec:extraction}

The Rosenbluth separation technique used in the extraction of $F_2$ and $F_L$ closely follows the method described in~\cite{Jimenez-Lopez:2024hpj}. 
 For $Q^2 \ll
 M_{Z}^2$ (mass of the Z boson squared), the inclusive cross section for Neutral Current (NC) DIS can be written in terms of $F_{2}$ and $F_{L}$ as

\begin{equation} \label{cross_sec}
    \dfrac{{\rm d}^2 \sigma^{e^{\pm}p}}{{\rm d}x {\rm d}Q^2} = \dfrac{2 \pi \alpha^2 Y_{+}}{xQ^4} \left[ F_{2}(x, Q^2) - \dfrac{y^2}{Y_{+}} F_{L}(x, Q^2) \right] = \dfrac{2 \pi \alpha^2 Y_{+}}{xQ^4} \sigma_{r}(x, Q^2, y) \ ,
\end{equation}

where $\alpha$ is the fine structure constant, 
$Y_{+} = 1 + (1 - y)^2$, and $\sigma_{r}$ is usually referred to as the reduced cross section. 

Eq.~\ref{cross_sec} implies a linear relationship between the reduced cross section and $y^2 / Y_{+}$, which is a function only of the inelasticity of the process, and is adjustable 
at fixed $x$ and $Q^2$
by changing the centre-of-mass energy and exploiting the relationship
$Q^2 \simeq s x y$.
Using measurements of $\sigma_{r}$ at different $s$,  the values of $F_{2}$ and $F_{L}$ can thus be separately obtained in a model independent way 
as the free parameters of a linear fit.
This Rosenbluth-type separation technique \cite{PhysRev.79.615}
has been applied to fixed target data 
and at HERA,
and has also recently been 
used in a simulated extraction of the diffractive longitudinal structure function at the EIC~\cite{PhysRevD.105.074006}.
For each $(x, Q^2)$, we thus apply a fit of the form
\begin{equation}\label{fit}
    \sigma_r(x,Q^2,y) = F_2(x,Q^2) - \frac{y^2}{Y_+} F_L(x, Q^2) \ ,
\end{equation}
where $F_2(x,Q^2)$ and $F_L(x,Q^2)$ are free parameters, obtained
through a $\chi^2$ minimisation procedure as
implemented in PYTHON using SciPy \cite{Virtanen:2019joe},
with the uncertainties being obtained from 
variations of
each parameter such that $\Delta \chi^2 = 1$. 

The pseudo-data smearing procedure introduces randomness into the fit inputs, which is reflected in the results as 
fluctuations in both the
values and uncertainties of the extracted structure functions.
The $F_2$ and $F_L$ uncertainty results are therefore samples from distributions of
possible outcomes which are often quite broad.
In order to sample the distribution of possible uncertainty outcomes
in a systematic way, we adopt the method 
introduced in~\cite{PhysRevD.105.074006} and 
used in~\cite{Jimenez-Lopez:2024hpj},
whereby the results are averaged over multiple replicas of the procedure.
With a sufficiently large number of replicas
(1000 is used by default), both the mean 
and the variance 
tend towards their
expectation values for the simulated scenario. 
The average and variance are calculated as:

\begin{equation}\label{avg_proc_1}
    \overline{v} = \frac{S_{1}}{N}, \quad (\Delta v)^2 = \frac{S_{2} - S^2_{1}/N}{N - 1},
\end{equation} 
where \( S_{n} = \sum_{i = 1}^{N} v^n_{i} \) and \( v_{i} \) is the value of structure function in the \( i \)-th MC sample. 

\subsection{Structure function results from combining HERA and two EIC beam configurations}

\begin{figure}[htb]
    \centering
    \includegraphics[origin=c,scale = 0.45]{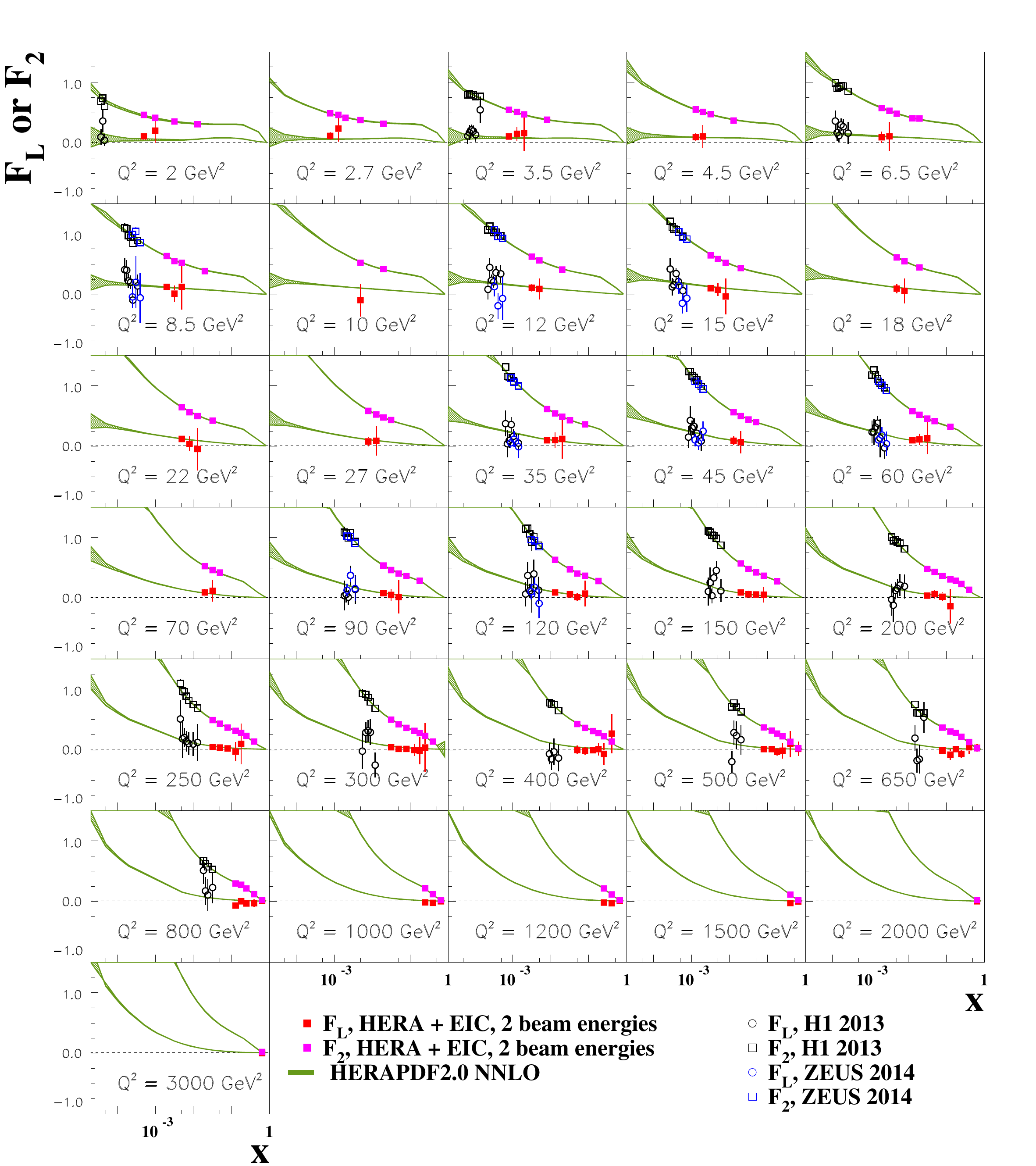}
    \caption{Simulated extractions of $F_2$ and $F_{L}$ averaged over $1000$ replicas for the 
    fits with the HERA and EIC data, for two EIC beam energy configurations.
    (The two EIC centre-of-mass energies are $100$ and $72$ GeV.)
    The error bars on the points represent
    the total experimental uncertainties. For the HERA+EIC $F_L$ 
    measurements, points with absolute uncertainties larger than 0.5 are removed for visual clarity but all points for the $F_2$ measurements are shown.}     
    \label{flx2}
\end{figure}

We first study a potential extraction 
in which EIC pseudo-data 
at the two currently planned $\sqrt{s}$ values
are fitted together with HERA inclusive data at four different centre-of-mass energies, such that there are a total of six distinct beam energy configurations.
Example results are shown for the $F_2$ and $F_L$ structure functions and their uncertainties as obtained using this method in 
Fig.~\ref{flx2}.
For the HERA+EIC $F_L$ measurements, points with absolute uncertainties larger than 0.5 are removed for visual clarity but all points for the $F_2$ measurements are shown.
The HERA results 
and predictions from HERAPDF2.0 NNLO are 
also shown. For a given $Q^2$ bin, the HERA 
$F_L$ measurements span a limited lever arm in $x$. 
With the inclusion of the EIC pseudo-data, 
model independent measurements 
of $F_2$ and $F_L$ 
may be obtained for larger values of $x$, 
extending the 
$F_L$ range by up to two orders of magnitude.

\subsection{Structure function results from two EIC beam configurations alone}
\label{two-method}

The extraction of the $F_2$ and $F_L$ structure functions using the Rosenbluth method is 
usually based on three or more measurements at 
different centre-of-mass energies. 
The standard Rosenbluth separation procedure that was used in \cite{PhysRevD.105.074006} and employed for other scenarios considered in this paper
thus cannot be directly applied for the
baseline assumption of two beam energy configurations in 
early EIC science. However, a simplified approach can be
applied in which 
the slope and intercept of the line passing through the two
measurements of $\sigma_r$ when plotted against $y^2/Y_+$
are taken as directly corresponding to $F_L$ and $F_2$, respectively.
The uncertainties on the structure functions 
can then be calculated using
standard error propagation.

\begin{figure}[htb]
    \centering
    \includegraphics[origin=c,scale = 0.45]{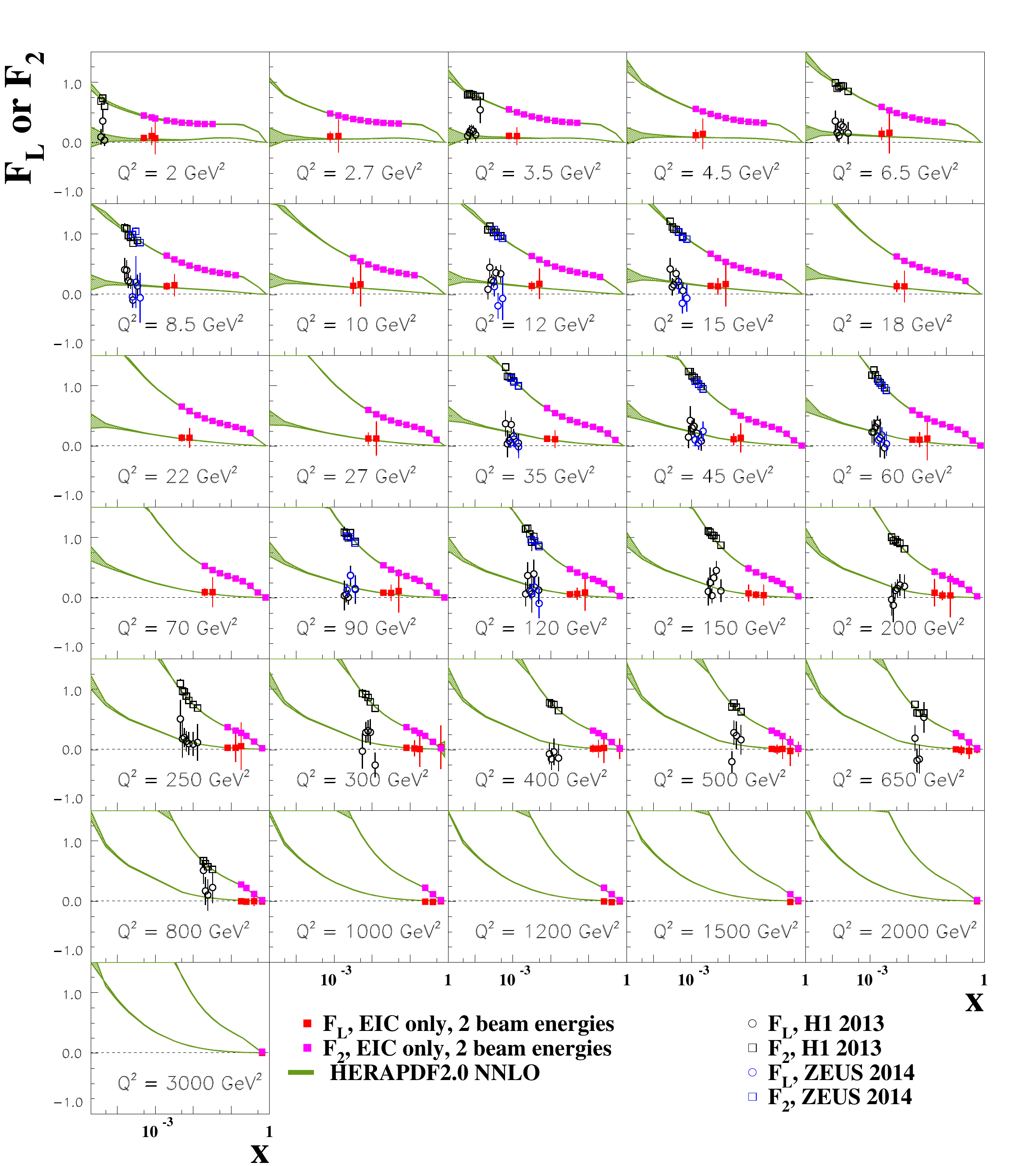}
    \caption{Simulated extractions of $F_2$ and $F_{L}$ averaged over $1000$ replicas for the 
    fits with  only two EIC beam energy configurations. 
    (The two EIC centre-of-mass energies are $100$ and $72$ GeV.)
    The error bars on the points represent
    the total experimental uncertainties. For the EIC $F_L$ 
    measurements, points with absolute uncertainties larger than 0.5 are removed for visual clarity but all points for the $F_2$ measurements are shown.}     
    \label{flx3}
\end{figure}

The expectation values and uncertainties of $F_2$ and $F_L$, obtained using only the two early science EIC beam energy configurations, are shown in Fig.~\ref{flx3}.
There are $\sim\SI{15}{\percent}$ fewer $F_L$ points with uncertainties of better than $0.5$ (as chosen for plotting) in the EIC-only scenario compared to the HERA and EIC combined scenario. 
The impact of adding of a third EIC 
beam energy configuration, yielding a further $\sqrt{s}$ value
and a return to the $\chi^2$ minimisation approach, is investigated
in sections~\ref{flsec} and~\ref{f2sec}.

\subsection{Kinematic coverage and precision for $F_L$ with two or three EIC beam configurations}
\label{flsec}

\begin{figure}[htb]
    \centering
     \includegraphics[scale = 0.425]{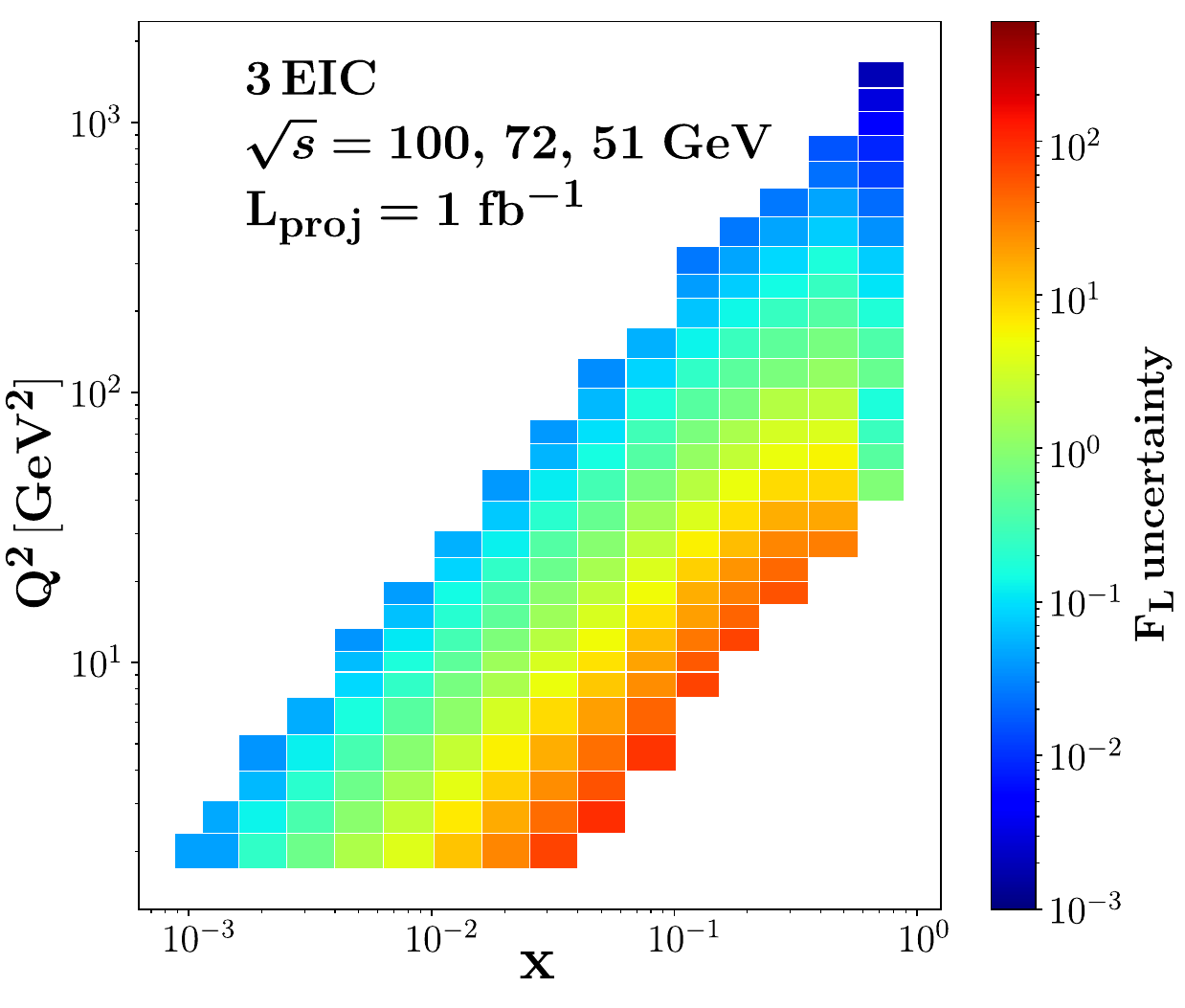} 
     \includegraphics[scale = 0.425]{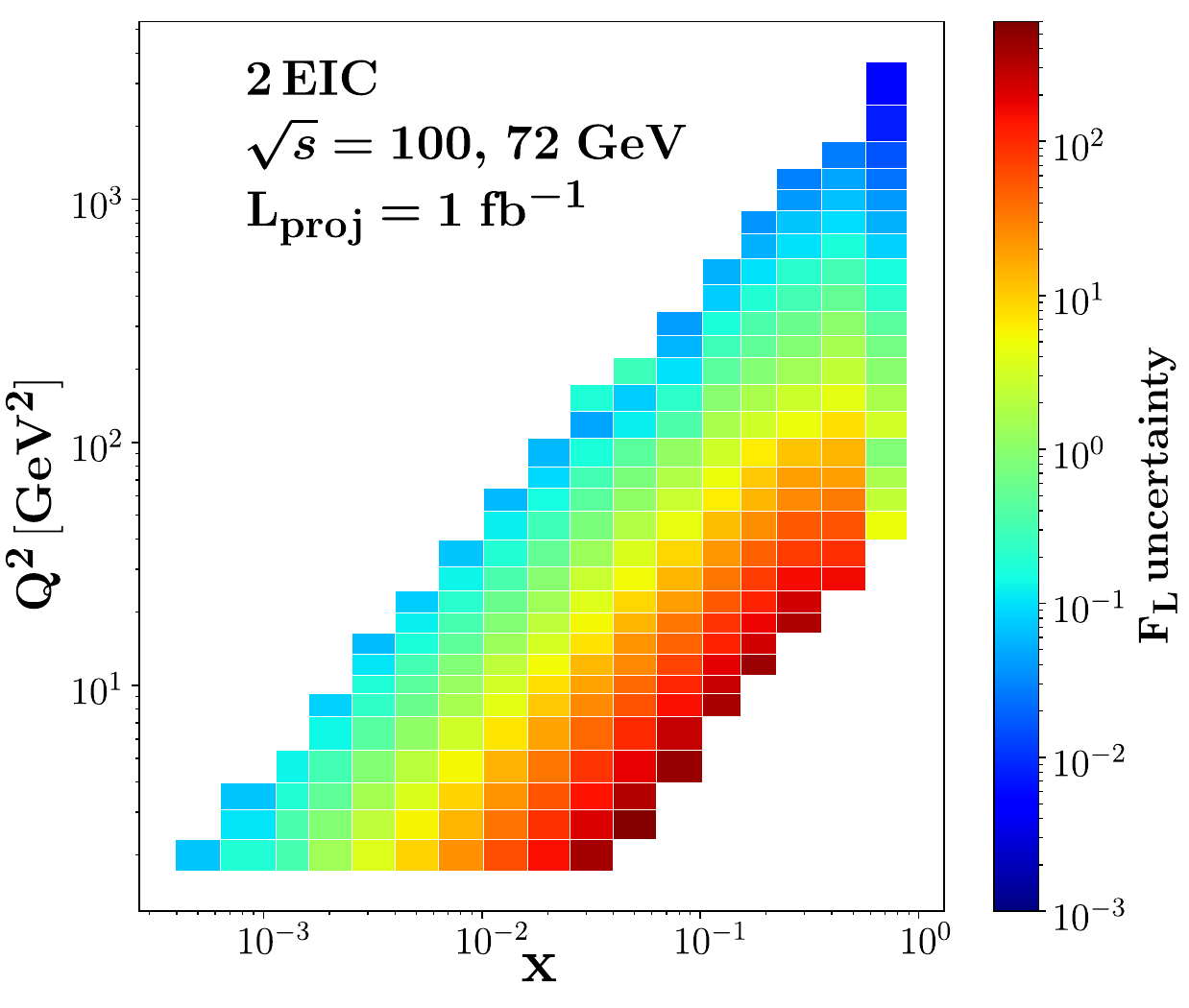} 
    \caption{Absolute uncertainties on the simulated EIC-only $F_L$ measurements averaged over 1000 replicas, corresponding to
the three (left) and two (right) beam energy configurations, with the colours indicating the uncertainties. 
Note that different methods are applied in the two cases, with 
all three beam energies required to be available in the fits for the `3 EIC' extraction and just two required for the `2 EIC' case.
}
    \label{hm2}
\end{figure}

\begin{figure}[htb]
    \centering
     \includegraphics[scale = 0.5]{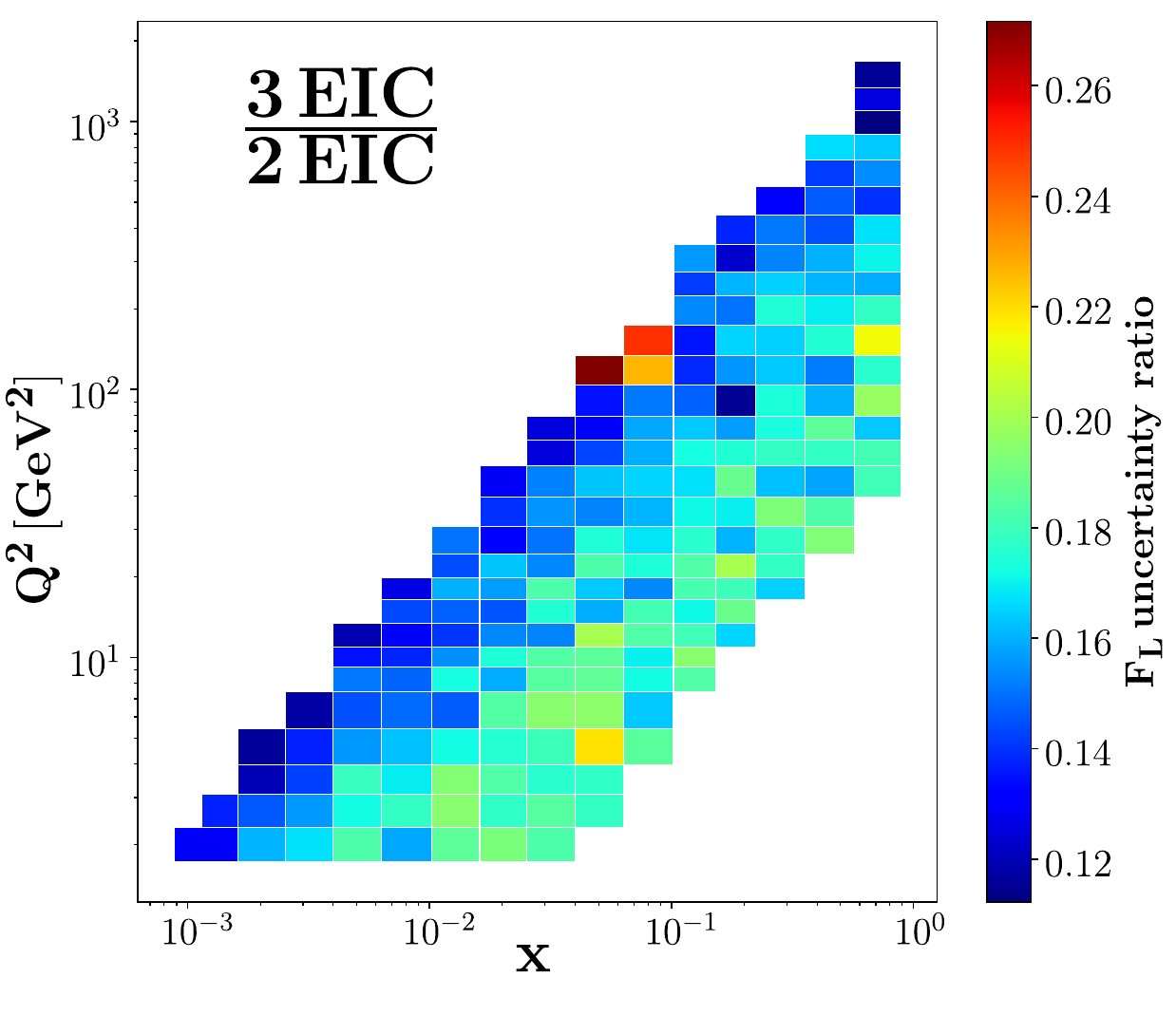}  
    \caption{Ratio of absolute uncertainties on the simulated EIC $F_L$ measurements for the three beam energy configuration 
    over the two beam energy configuration, averaged over 1000 replicas.
    For the three beam energy configuration the EIC centre-of-mass energies are $100$, $72$ and $51$ GeV. 
    For the two beam energy configuration the EIC centre-of-mass energies are $100$ and $72$ GeV.
    All three beam energies are required in the fits for the `3 EIC' extraction, whereas just two are 
    required for the `2 EIC' case.
}     
    \label{hm1_ratio}
\end{figure}

So far, we have considered only two distinct EIC $ep$ beam energy settings, achieved via the scattering of \SI{10}{\giga\electronvolt} electrons on \SI{130}{\giga\electronvolt} or \SI{250}{\giga\electronvolt} protons. 
Here we additionally
present results when the third, lowest, EIC centre-of-mass energy,
resulting from scattering of  \SI{5}{\giga\electronvolt} electrons with \SI{130}{\giga\electronvolt} protons, is included in the analysis.
Figure~\ref{hm2} shows absolute uncertainties on the simulated $F_L$ measurements for the EIC-only scenarios, 
corresponding to two or three beam energy configurations, averaged over 1000 replicas. 
The difference in extraction methods and the number of data points used in the fits leads to 
very different uncertainties for the two and three-beam $F_L$ results. 
This can be seen in Fig.~\ref{hm1_ratio} which shows the ratio of the absolute 
uncertainties for the three- to two-beam $F_L$ results in the common kinematic region. 
The fits with the EIC-only data and three beam configurations give significantly more precise results for the $F_L$ structure function compared 
to the two-beam scenario, the ratio typically falling between 
10\% and 20\%  and never exceeding 32\%.


\begin{figure}[htb]
    \centering
     \includegraphics[scale = 0.425]{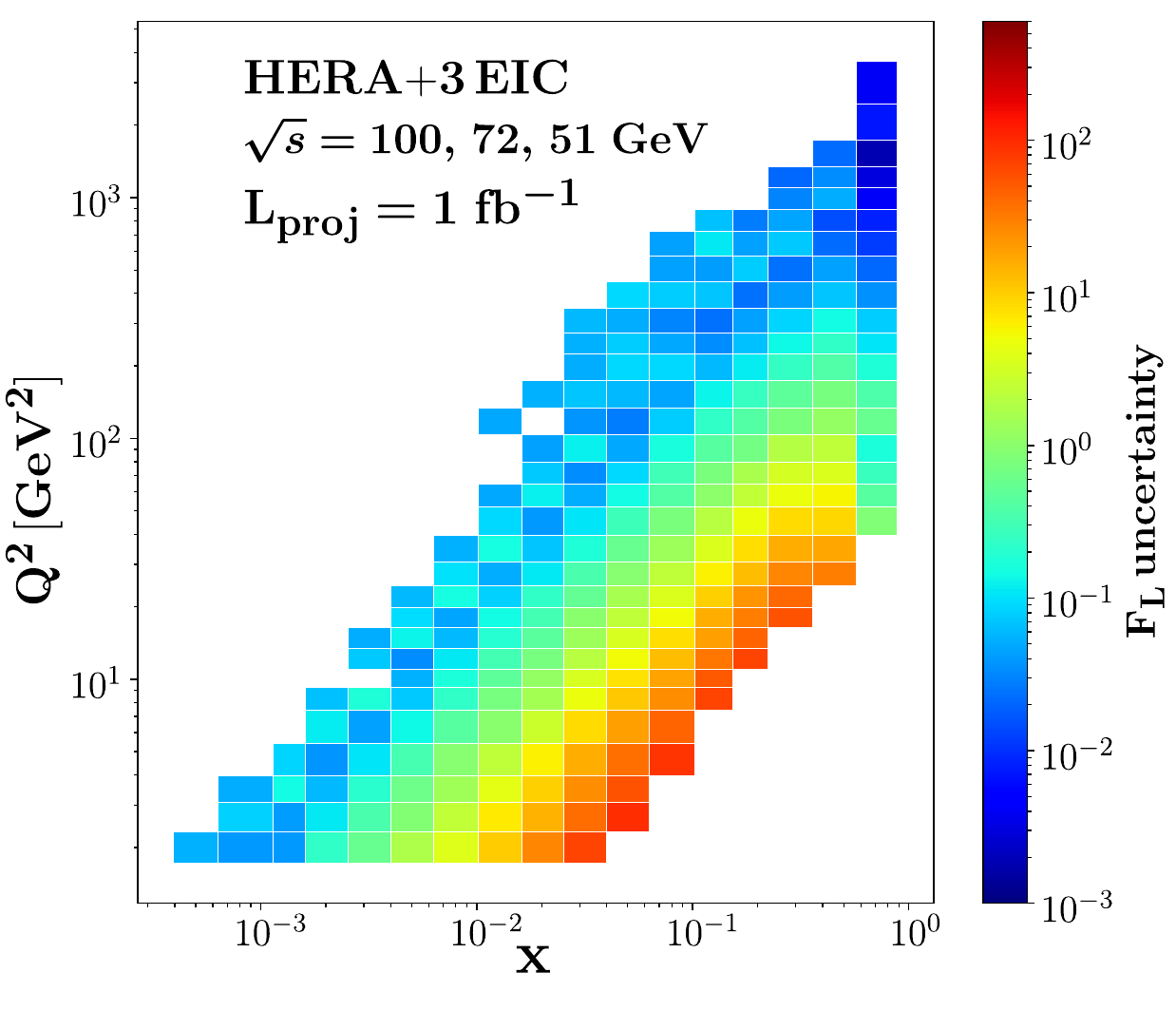} 
     \includegraphics[scale = 0.425]{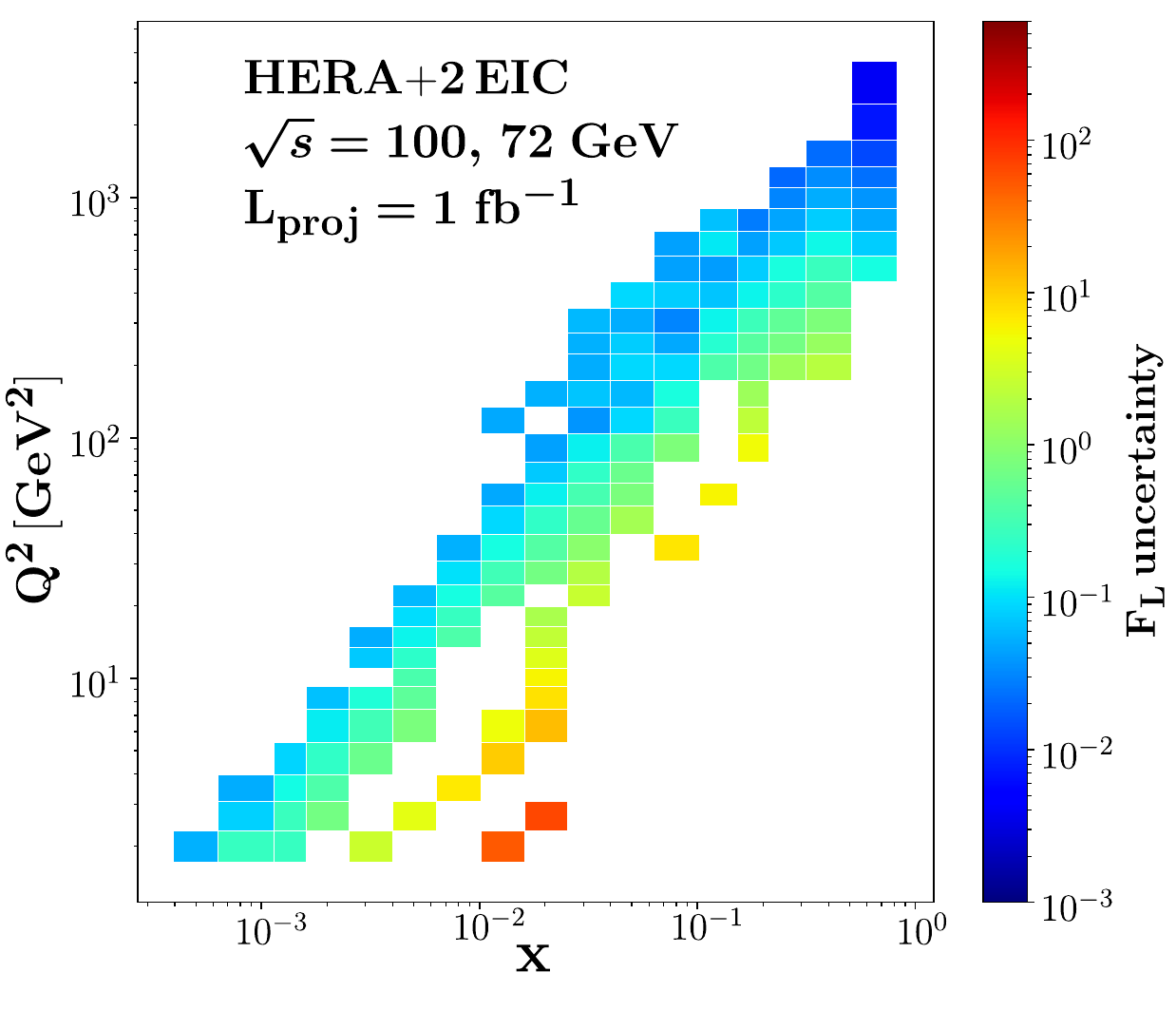} 
    \caption{Absolute uncertainties on the simulated HERA+EIC $F_L$ measurements averaged over 1000 replicas, corresponding to
the three (left) and two (right) beam energy configurations, with the colours indicating the uncertainties.
At least three data points are required for the Rosenbluth
fits in both cases. 
}     
    \label{hm1}
\end{figure}

\begin{figure}[htb]
    \centering
     \includegraphics[scale = 0.5]{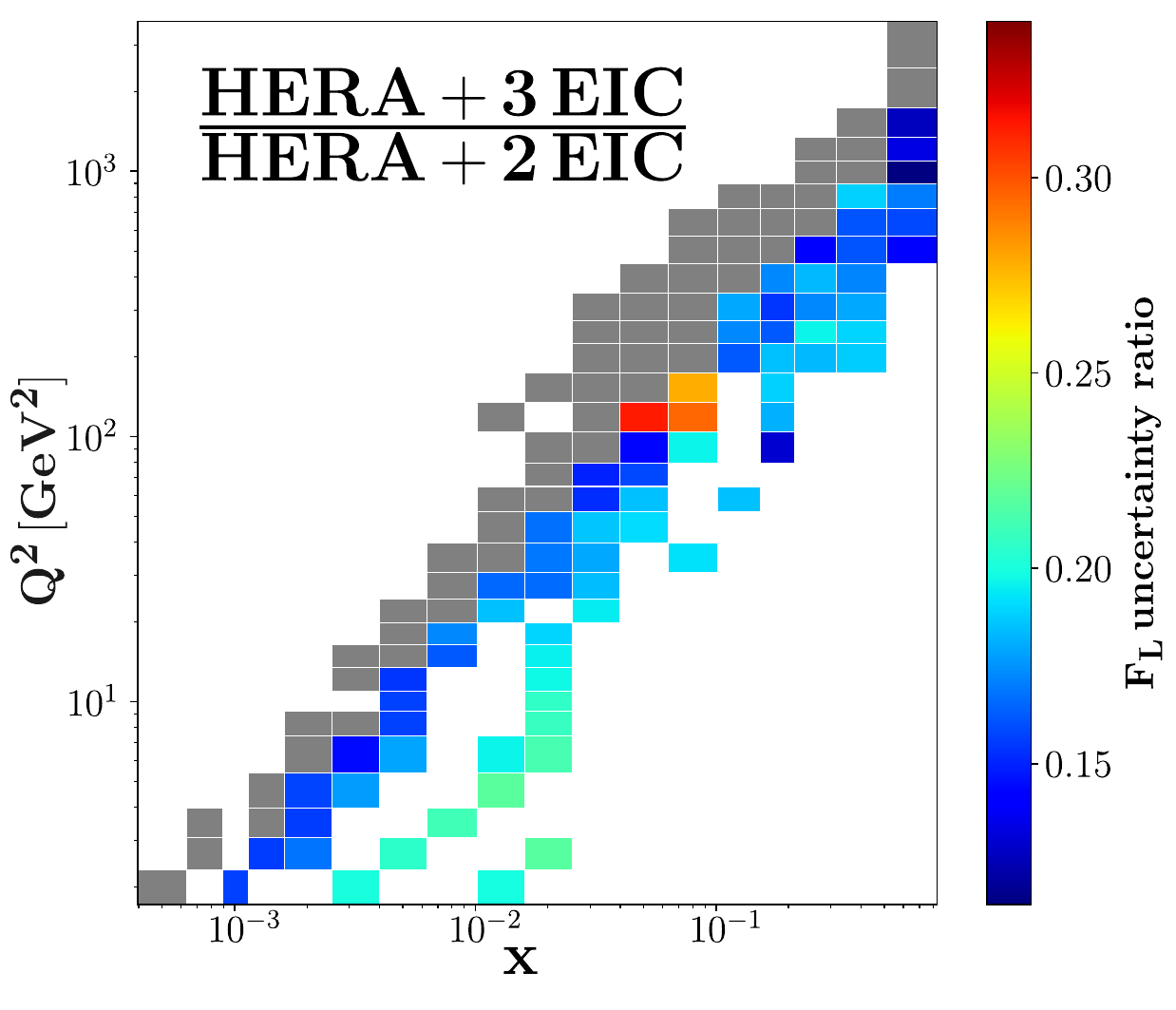}  
    \caption{Ratio of absolute uncertainties on the simulated EIC $F_L$ measurements for the three beam energy configuration 
    over the two beam energy configuration, averaged over 1000 replicas.
    For the three beam energy configuration the EIC centre-of-mass energies are $100$, $72$ and $51$ GeV. 
    For the two beam energy configuration the EIC centre-of-mass energies are $100$ and $72$ GeV.
At least three data points are required for the Rosenbluth
fits in both cases. 
The grey boxes indicate cases where the ratio is exactly 1
by construction, because the same number of EIC points 
enter the Rosenbluth fits for the two- and three-beam scenarios.
}     
    \label{hm1_ratio2}
\end{figure}

Figure~\ref{hm1} shows absolute uncertainties on the simulated $F_L$ measurements
for HERA data combined with two or three EIC beam energy configurations, averaged over 1000 replicas. 
Here, for both two and three EIC beam energies, at least three data points are required to perform the Rosenbluth fits. 
As a result, the phase-space for measurements using HERA and two EIC beam energies is considerably smaller than 
that for HERA and three EIC beam energies. \textcolor{black}{Gaps in coverage appear in Fig.~\ref{hm1_ratio2} where no HERA measurement is available for the chosen bin, due the the coarser binning of HERA measurements at large $x$ and low $Q^2$.} Measurements in the three-beam scenario are more precise, though the difference is not as large as for the EIC-only case.
This can be seen in Fig.~\ref{hm1_ratio2} which shows the ratio of the absolute 
uncertainties for the three- to two-beam $F_L$ results for the fits with HERA + EIC data in the common kinematic region. 
The fits with the HERA and EIC data and three EIC beam configurations give significantly more precise results for the $F_L$ structure function compared 
to the two-beam scenario, the ratio typically falling between 
10\% and 20\%  and never exceeding 30\%. \textcolor{black}{The precision of the $F_L$ extraction depends on the $y^2/Y_+$ lever arm at a given $x$ and $Q^2$, which increases roughly as $y_\text{max}^2$, where $y_\text{max}$ is the largest $y$ value included in the fit. With the inclusion of the lower energy EIC beam configuration, an additional high $y$ data point becomes available, extending the $y^2/Y_+$ lever arm by a factor of $\sim5-6$, which is reflected in the $F_L$ uncertainties.}


\begin{figure}[htb]
    \centering
     \includegraphics[scale = 0.45]{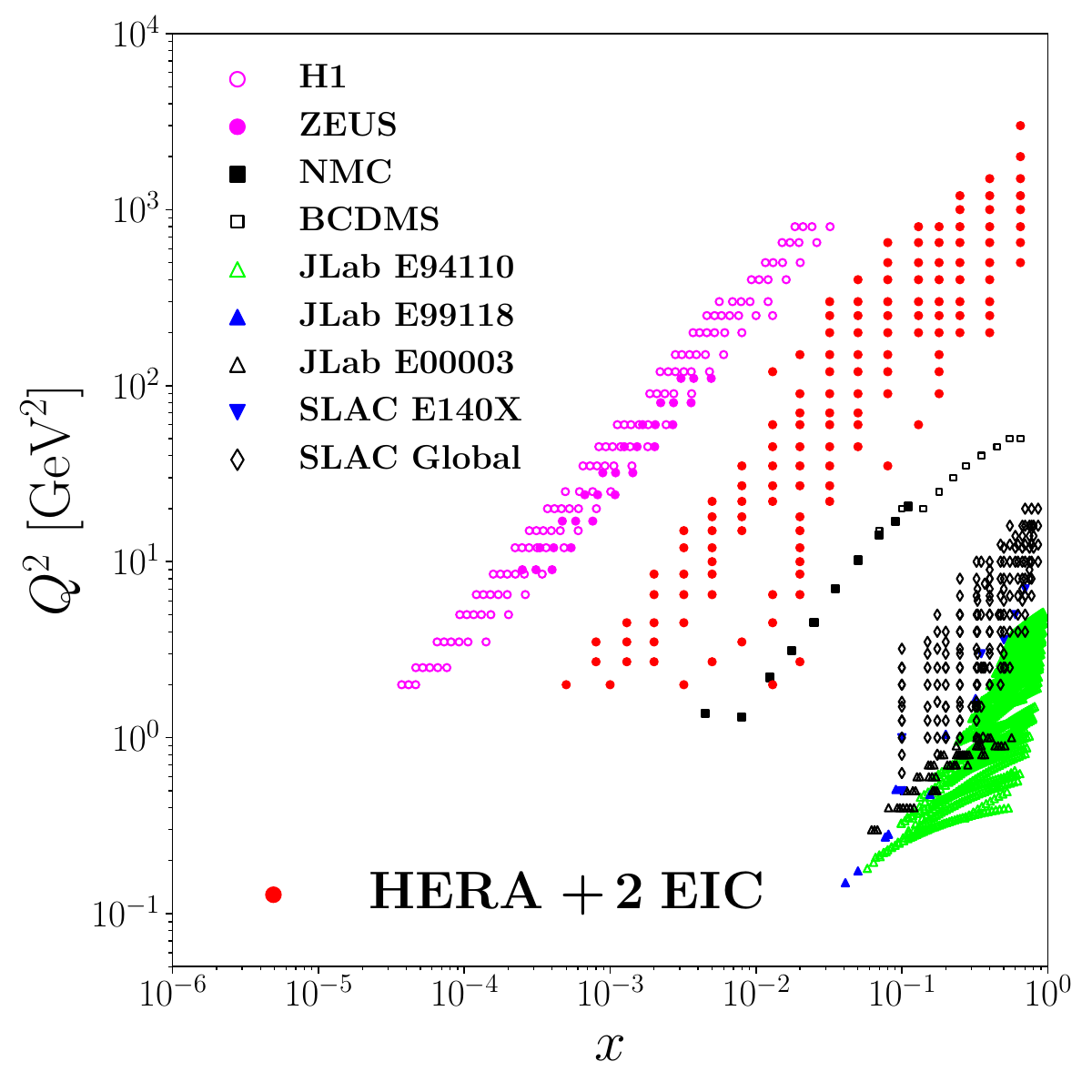}
     \includegraphics[scale = 0.45]{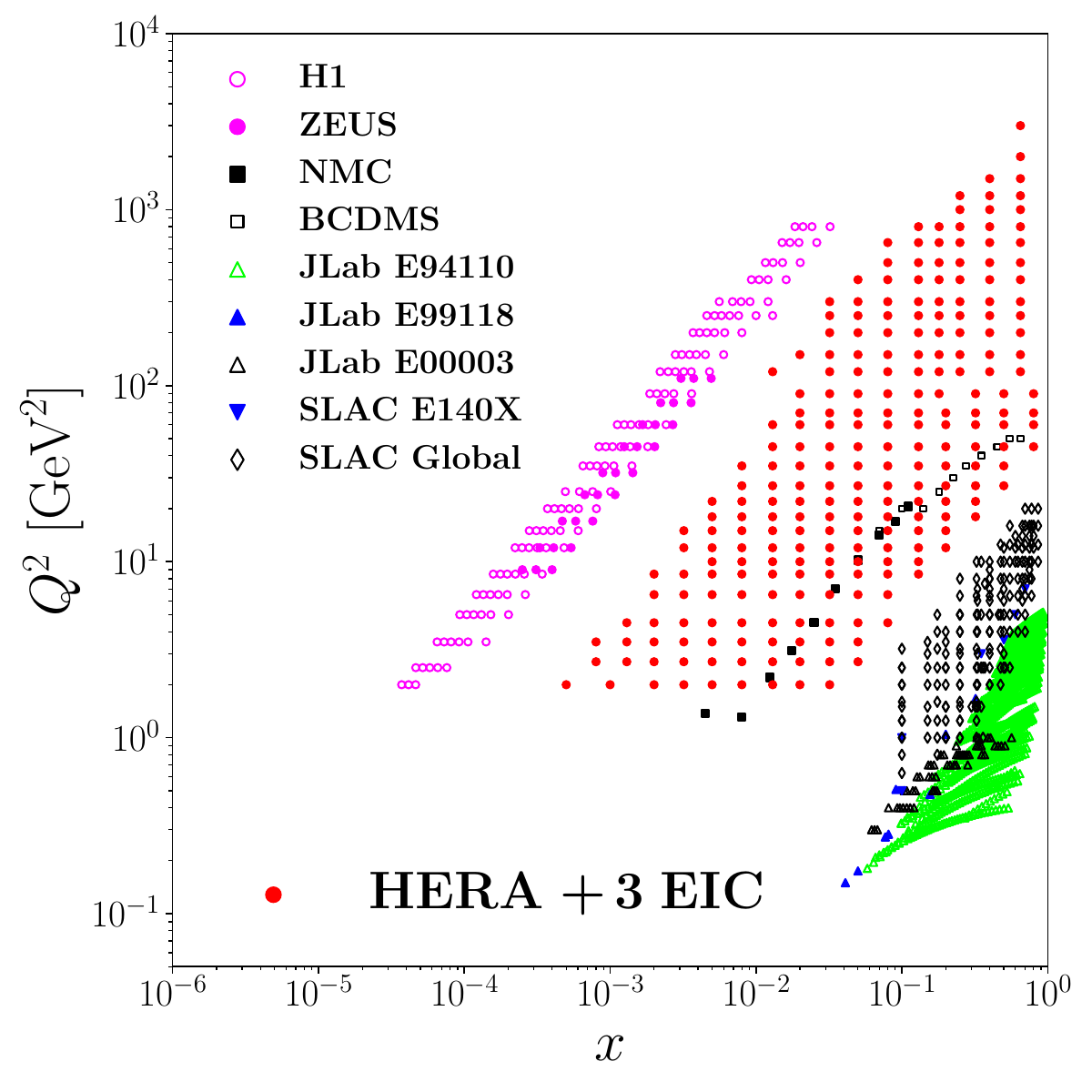} \\
     \includegraphics[scale = 0.45]{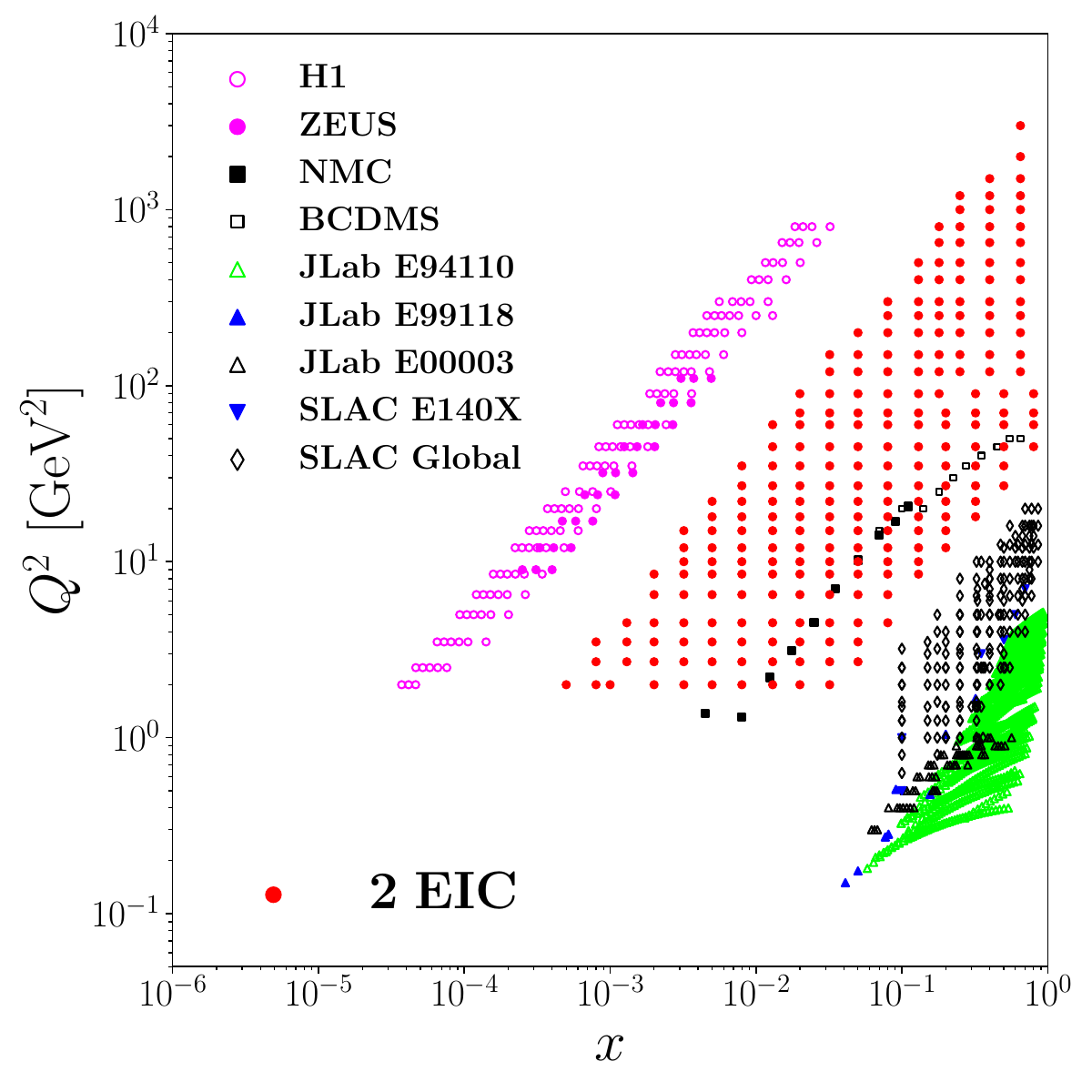}
     \includegraphics[scale = 0.45]{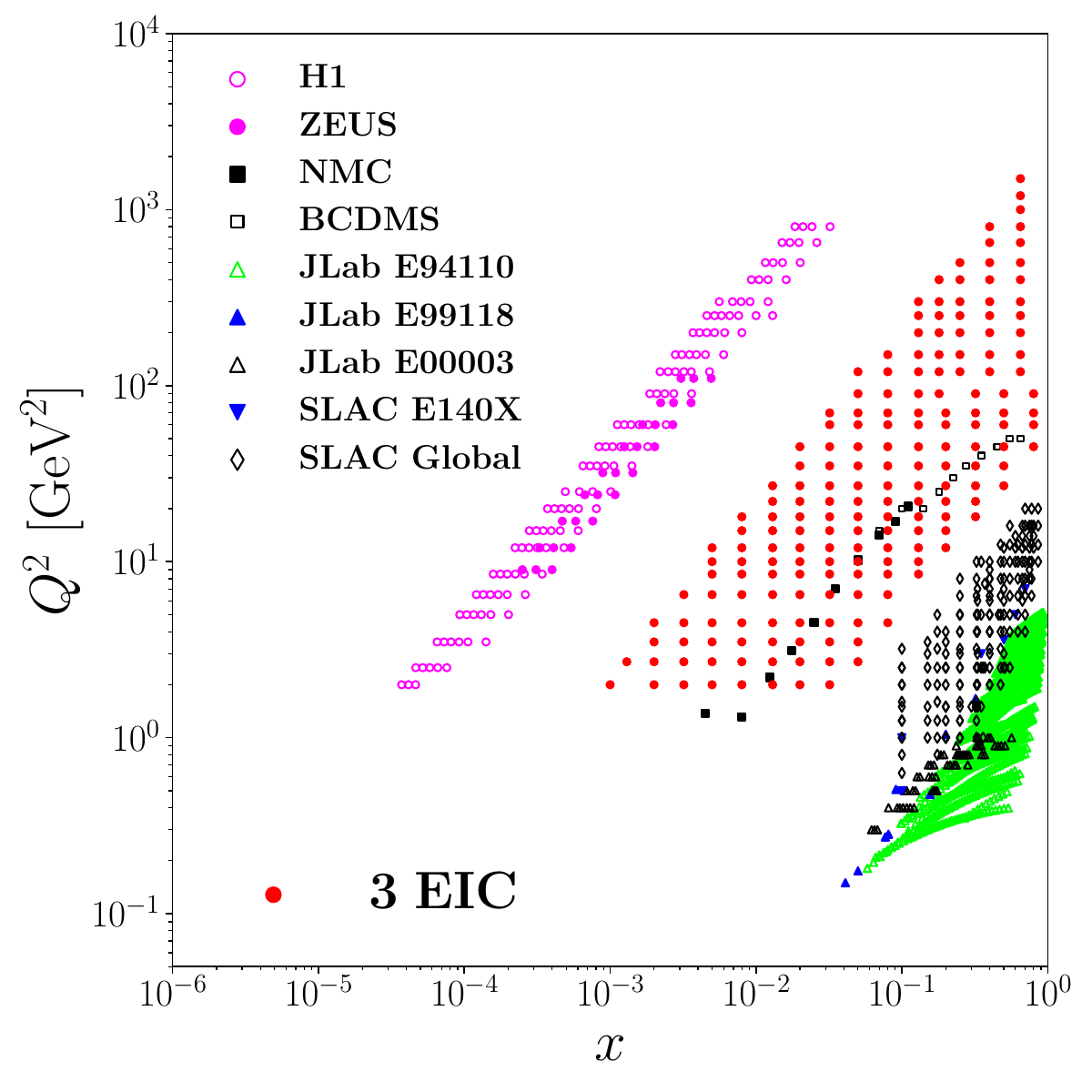}
    \caption{Kinematic coverage of world data for the proton longitudinal structure function, derived from figure 1 in
    \cite{PhysRevC.97.045204}, but with simulated EIC data added.
    The existing data shown are from H1 and ZEUS~\cite{H1:2013ktq,ZEUS:2014thn}, NMC~\cite{ARNEODO19973}, BCDMS~\cite{BENVENUTI1989485}, JLab~\cite{PhysRevC.97.045204,Tvaskis:2006tv,JeffersonLabHallCE94-110:2004nsn} and SLAC~\cite{E140X:1995ims,Whitlow:1991uw}.
    The EIC pseudo-data are shown for the fits with HERA + EIC data (top), and EIC data only (bottom), for the 
    two (left) and three (right) beam scenarios. 
    The EIC centre of mass energies are $\sqrt{s}=100$ and $72$ GeV for the two-beam case, with $\sqrt{s}=51$ GeV added for the three-beam case.
    At least three $\sqrt{s}$ values are required in the 
    Rosenbluth fits in all cases except for the `2 EIC' 
    extraction, where both $\sqrt{s}$ values are required.
    }     
    \label{kplane1}
\end{figure}

Figure~\ref{kplane1} shows the phase-space accessible for the 
four measurement strategies discussed here, together with 
a selection of the previous 
structure function measurements. 

\subsection{Kinematic coverage and precision for $F_2$ with two
or three EIC beam configurations}
\label{f2sec}

\begin{figure}[htb]
    \centering
     \includegraphics[scale = 0.425]{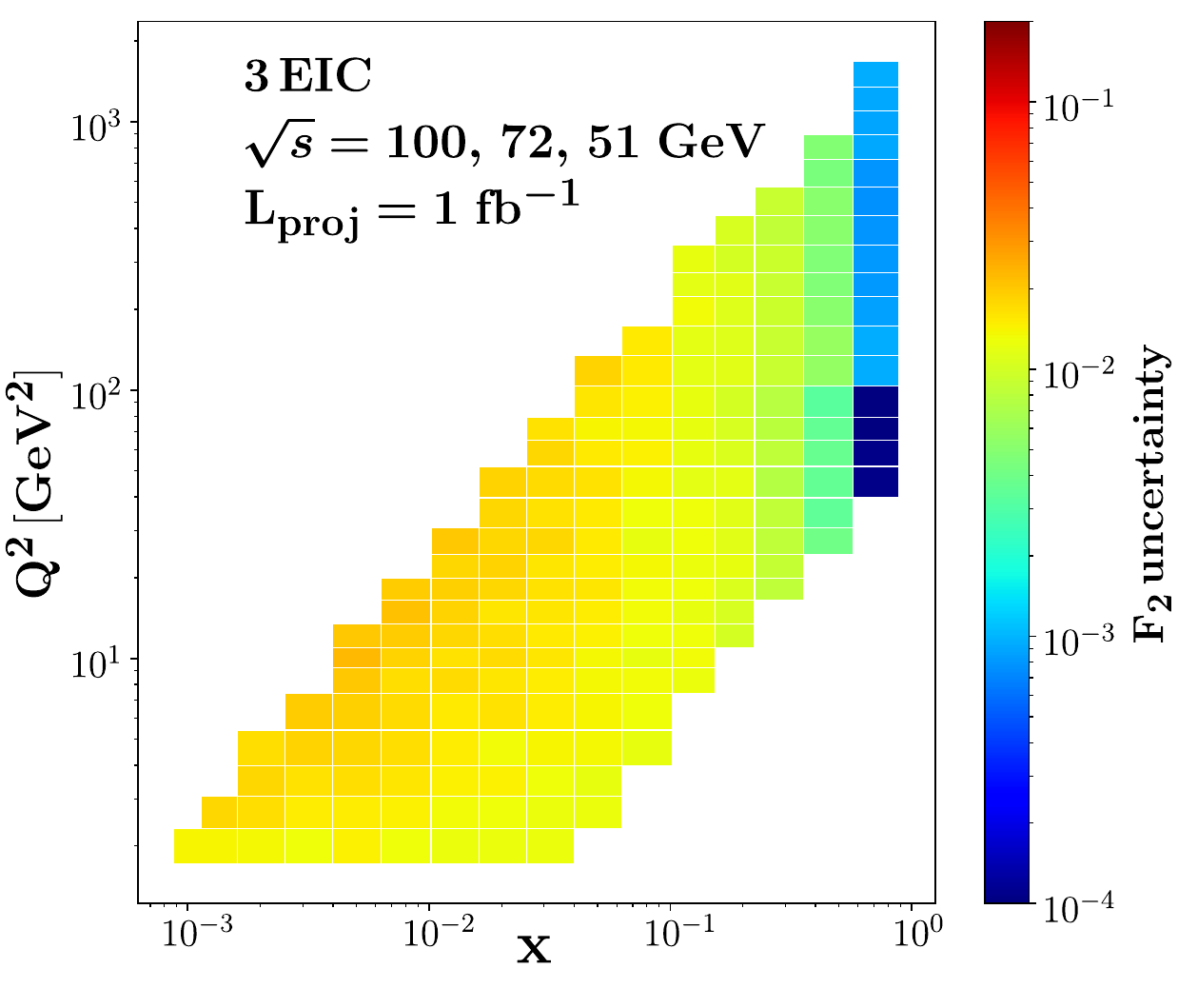} 
     \includegraphics[scale = 0.425]{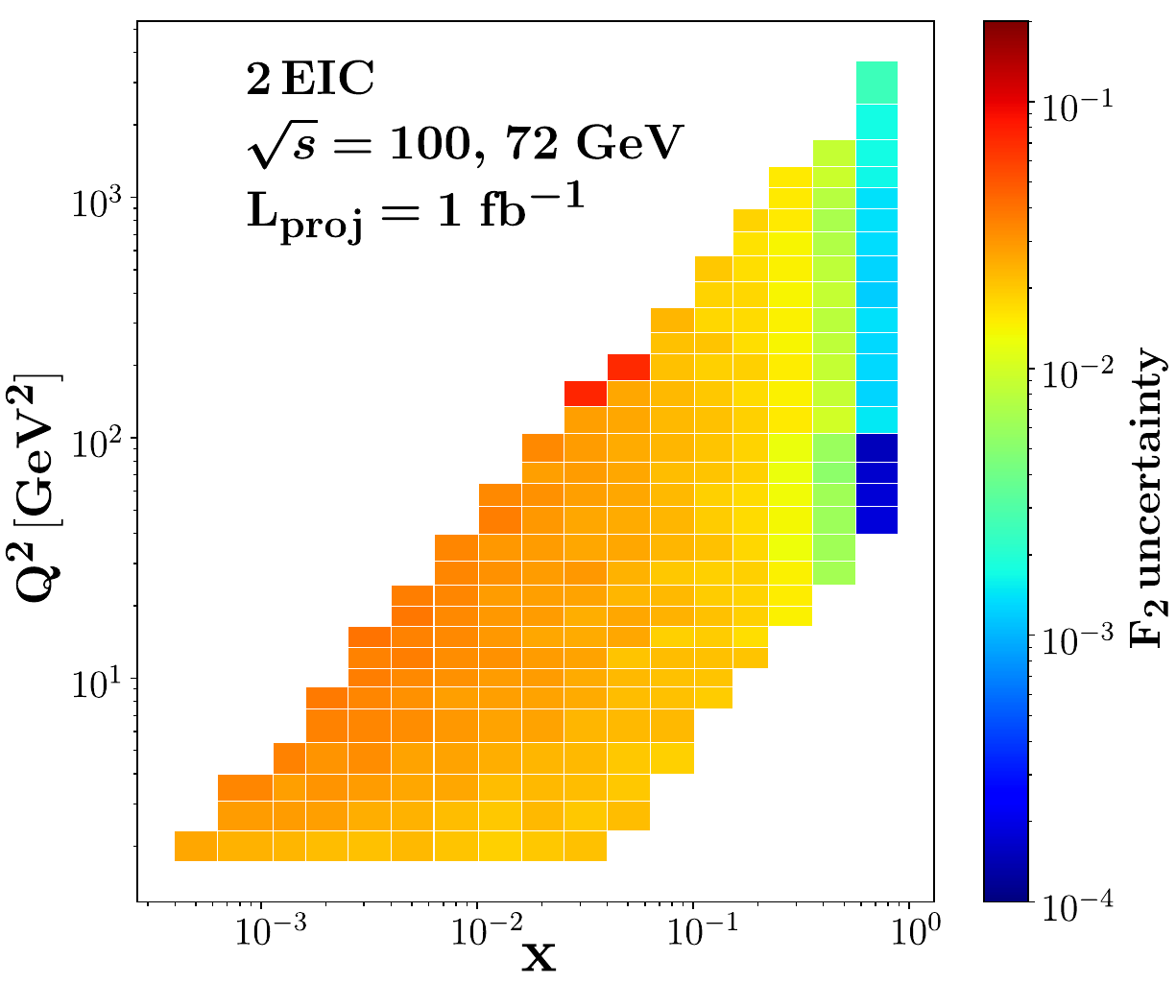} 
    \caption{Absolute uncertainties on the simulated EIC-only $F_2$ measurements averaged over 1000 replicas, corresponding to
the three (left) and two (right) beam energy configurations, with the colours indicating the absolute uncertainties.
}     
    \label{hm2_F2}
\end{figure}

\begin{figure}[htb]
    \centering
     \raisebox{0.85ex}{\includegraphics[scale = 0.425]{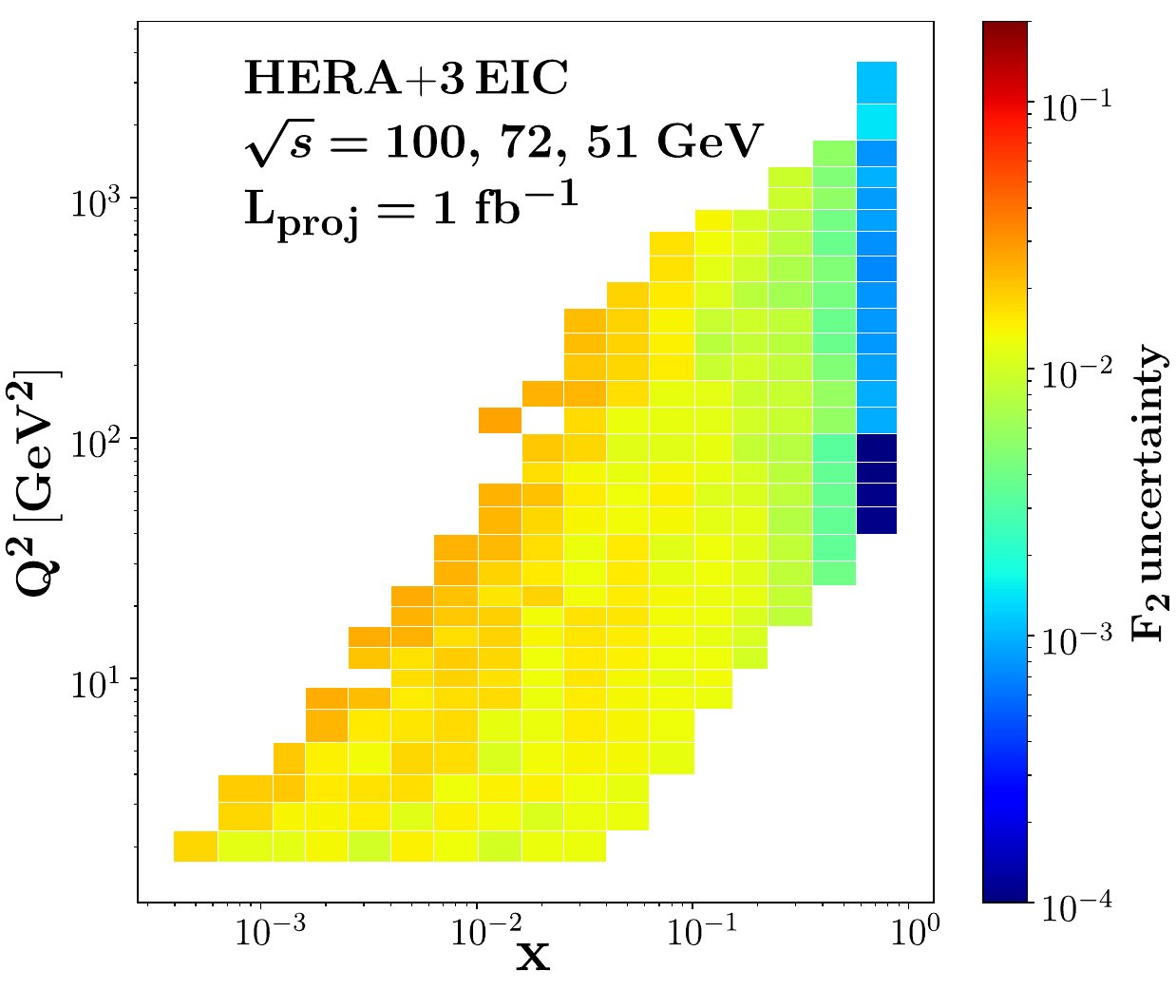}}
     \includegraphics[scale = 0.425]{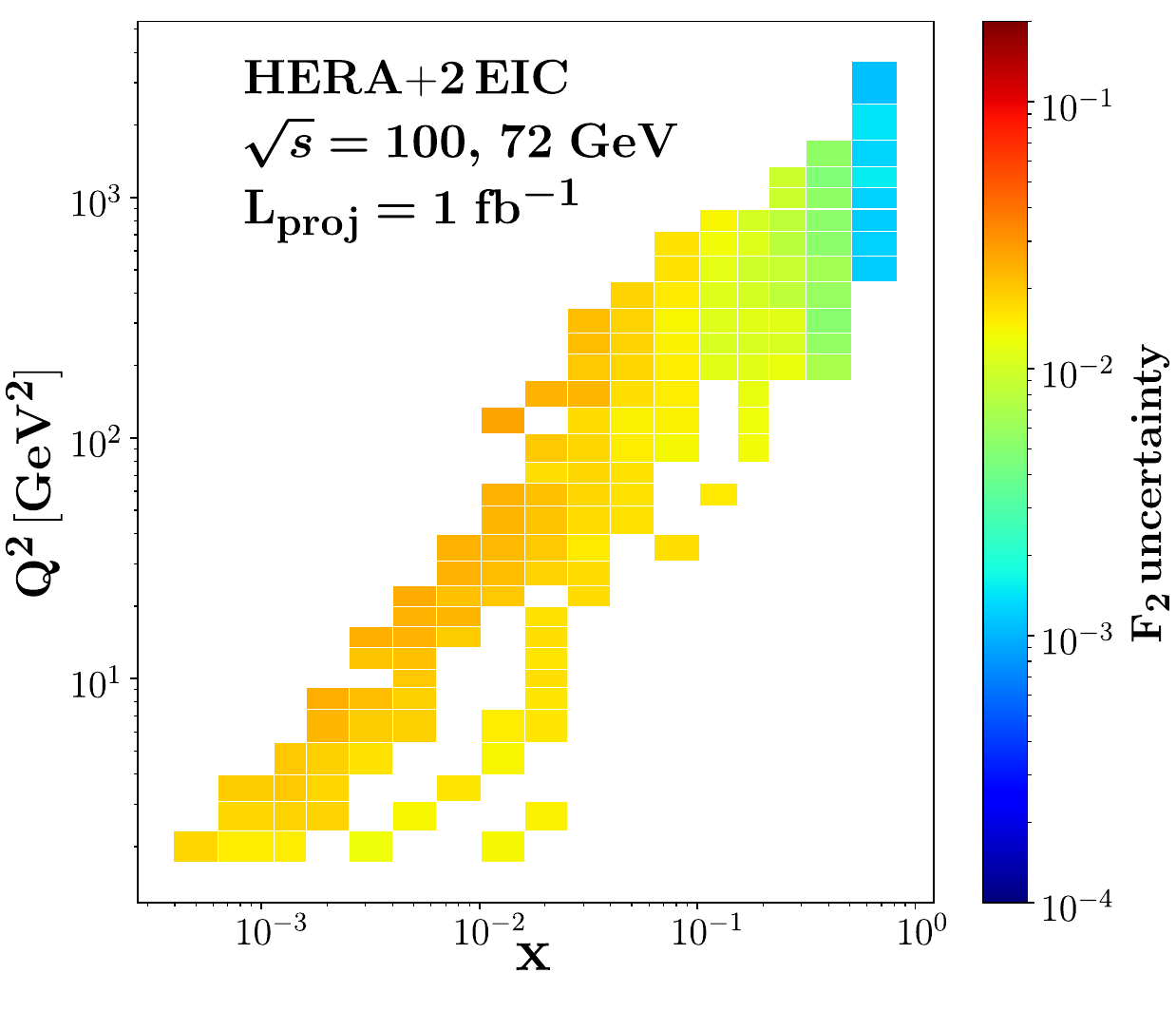} 
    \caption{Absolute uncertainties on the simulated HERA+EIC $F_2$ measurements averaged over 1000 replicas, corresponding to
the three (left) and two (right) beam energy configurations, with the colours indicating the absolute uncertainties.
}     
    \label{hm1_F2}
\end{figure}

Figures~\ref{hm2_F2} and~\ref{hm1_F2} show absolute uncertainties on the simulated $F_2$ extractions, 
for EIC-only and HERA+EIC 
simulations, respectively, 
averaged over 1000 replicas. 
The kinematic region for the various scenarios is exactly the same as for the $F_L$ extractions. 
The absolute uncertainties for 
$F_2$ are 
at the $10^{-2}$ level for most of the accessible kinematic plane. 
As expected, the 
three-beam scenario offers higher precision, for both EIC-only 
and HERA+EIC measurements.

In many HERA publications on inclusive cross sections when the $F_2$ structure function is extracted~\cite{H1:1999qgy,H1:2000olm,H1:2003xoe,H1:2009jxj,H1:2009bcq,ZEUS:2001mhd}, 
it is calculated from the 
reduced cross section $\sigma_r$ using  models to determine the 
size of the $F_L$ contribution.
In such publications, $F_2$ is either quoted without an uncertainty or the uncertainty is simply taken
as a relative total uncertainty of the cross section. 
No model uncertainties are calculated.
In contrast, the Rosenbluth separation method offers 
a completely model-independent extraction of $F_2$. 
Here we investigate the improvements over the HERA extractions
that can be obtained by adding EIC data and applying the 
Rosenbluth separation method. 
For all beam energies, the total uncertainties on the EIC
cross sections are taken as \SI{3.9}{\percent}  summed in quadrature with the statistical uncertainties (see Section~\ref{pseudo}).

\begin{figure}[htb]
    \centering
     \includegraphics[scale = 0.35]{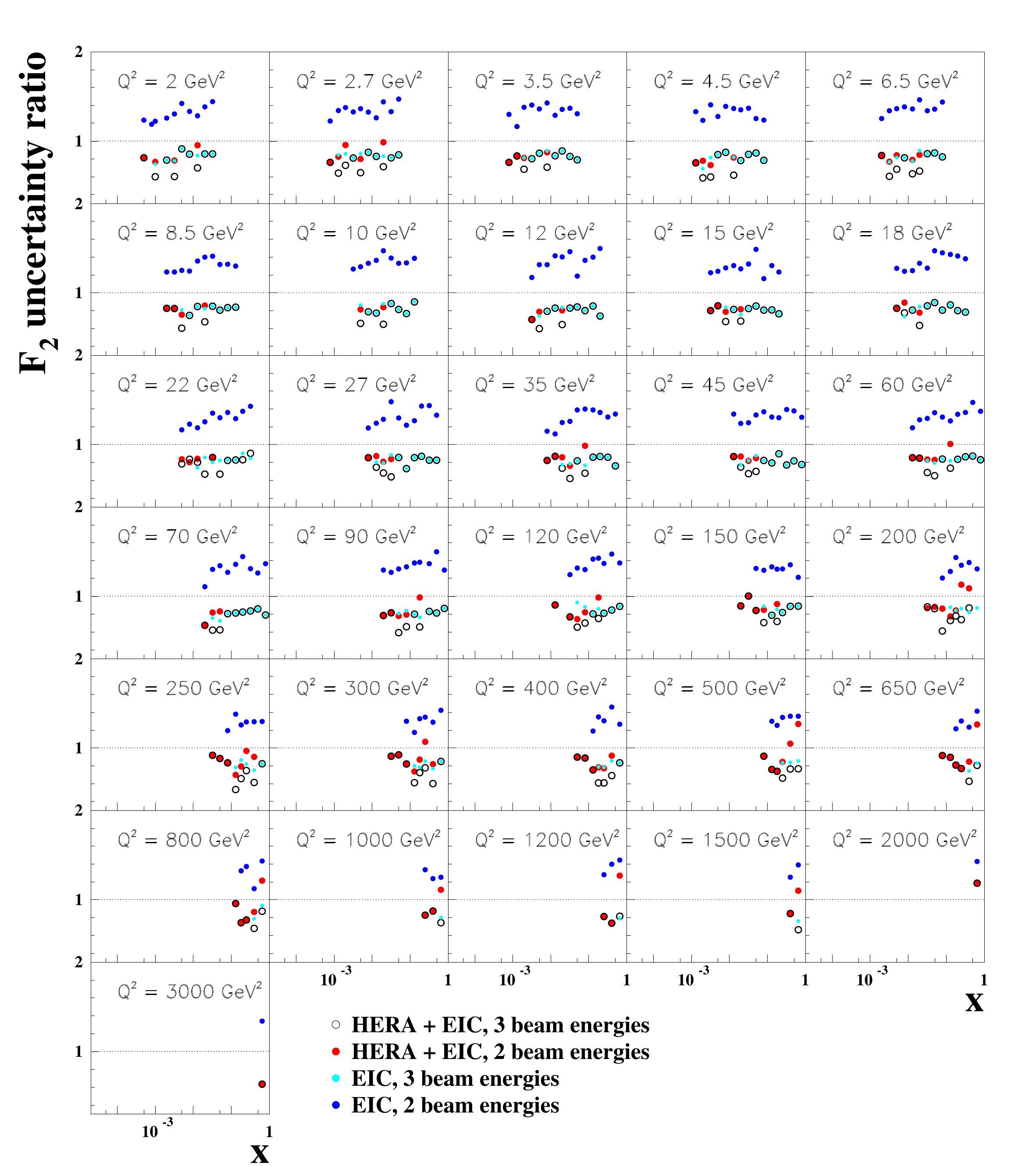} 
    \caption{The ratio of $F_2$ uncertainties obtained using the model-independent Rosenbluth   approach, to those obtained using HERA data only.
     Different symbols correspond to various beam energy combinations, for EIC-only and
     HERA+EIC fits.
     The beam settings are $\sqrt{s}=100$ and $72$ GeV for the two-beam case, with $\sqrt{s}=51$ GeV added for the three-beam case.
}     
    \label{f2-comp}
\end{figure}

Fig.~\ref{f2-comp} shows the ratio of $F_2$ uncertainties obtained using the model-independent Rosenbluth approach for 
various beam energy combinations, to those obtained using the
HERA data only
(i.e. with the fractional uncertainty taken directly 
from the reduced cross section). For the fits combining the HERA and EIC data and for the fits 
with EIC-only data with three centre-of-mass energies, the uncertainties obtained using the Rosenbluth separation 
are almost always lower than 
those from using HERA data alone.
Exceptions only occur in cases
where only one EIC point is used in the fit.
For the EIC-only fits with three beam energies, all $F_2$ measurements are 
more precise than the HERA results
in the regions of overlap, with improvements 
typically at the 20\% level. 
Conversely, due to the difference in method, 
the scenario with EIC-only data and only two energies 
always leads to less precise results than those from HERA.

\section{HERA+EIC collinear proton PDFs and strong coupling} 
\label{sec:pdfs}

Previous publications~\cite{armesto2024impact,Cerci:2023uhu} have shown that the full EIC DIS data will have 
a substantial impact on 
the precision of PDFs extracted in fits to DIS data only 
and on the strong coupling. 
Following these studies, similar analyses are performed here 
assuming
1~fb$^{-1}$ of early science EIC data for the two- and three-beam configuration scenarios.
The results presented in this section are obtained from global QCD fits at NNLO,
performed in the HERAPDF2.0 
framework~\cite{H1:2015ubc} using xFitter, 
an open source QCD fit platform~\cite{Alekhin:2014irh}.
Fits with identical configurations are performed
to HERA data only, corresponding to 
HERAPDF2.0 NNLO~\cite{H1:2015ubc}, and also 
with the additional inclusion of the simulated EIC early science NC DIS pseudo-data described in Section~\ref{pseudo}. 
To avoid regions that may be strongly affected by higher twist
or resummation effects, a cut on the squared 
hadronic final state invariant mass,
$W^2 = Q^2 (1-x) / x > 10$~GeV$^2$ is applied for all data. 
The central values of the PDFs with and without the EIC pseudo-data
coincide by construction, so the uncertainties can be compared directly.

\begin{figure}[htb]
    \centering
     \includegraphics[scale = .79]{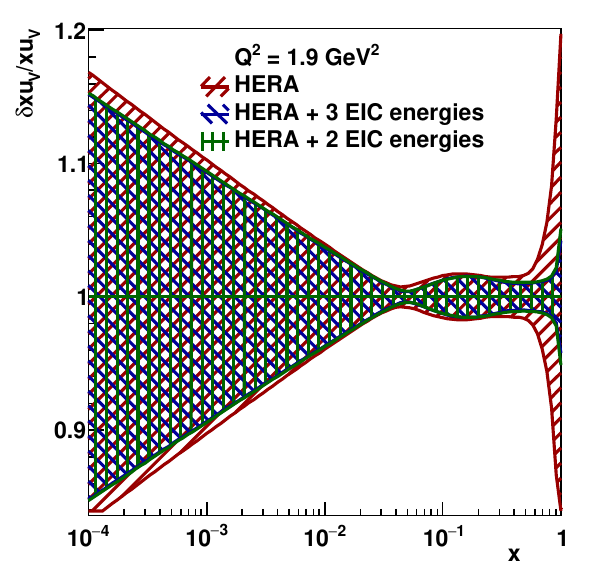}
     \hspace{0.05cm}
     \includegraphics[scale = 0.363]{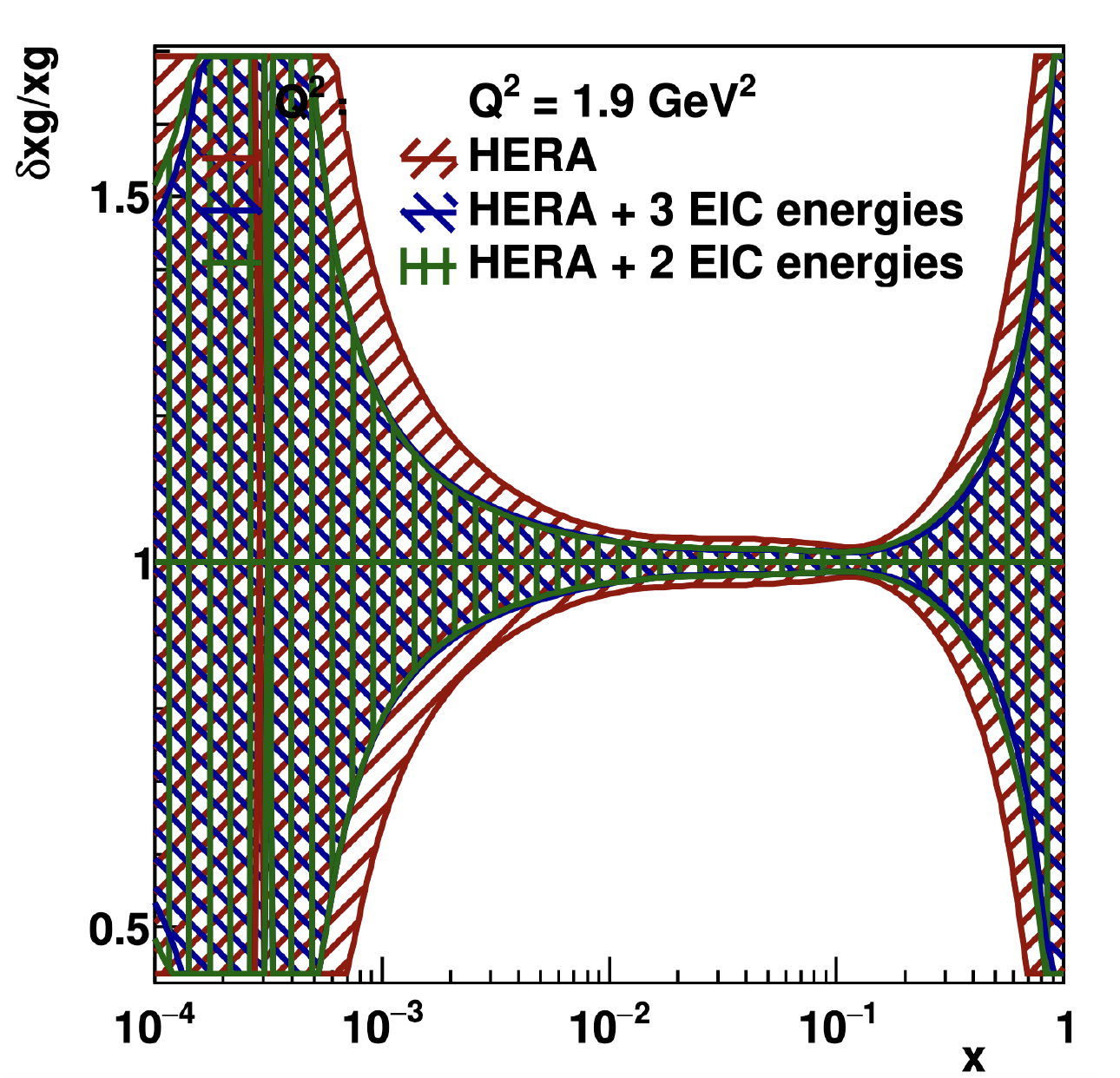}
     \caption{Impact of simulated EIC data on the NNLO collinear parton distributions of the proton (logarithmic $x$ scale). 
     The bands show relative experimental uncertainties for the up-valence and gluon distributions 
     for $Q^2 = \SI{1.9}{\giga\electronvolt\squared}$. The HERAPDF2.0NNLO total uncertainties (using HERA data alone) are compared with results 
     in which simulated EIC data for different beam 
     energy scenarios are also included in the HERAPDF2.0NNLO fitting framework.
     Beam settings are $\sqrt{s}=100$ and $72$ GeV for the two-beam case, with $\sqrt{s}=51$ GeV added for the three-beam case.}
    \label{pdf1}
\end{figure}

\begin{figure}[htb]
    \centering
     \includegraphics[scale = .36]{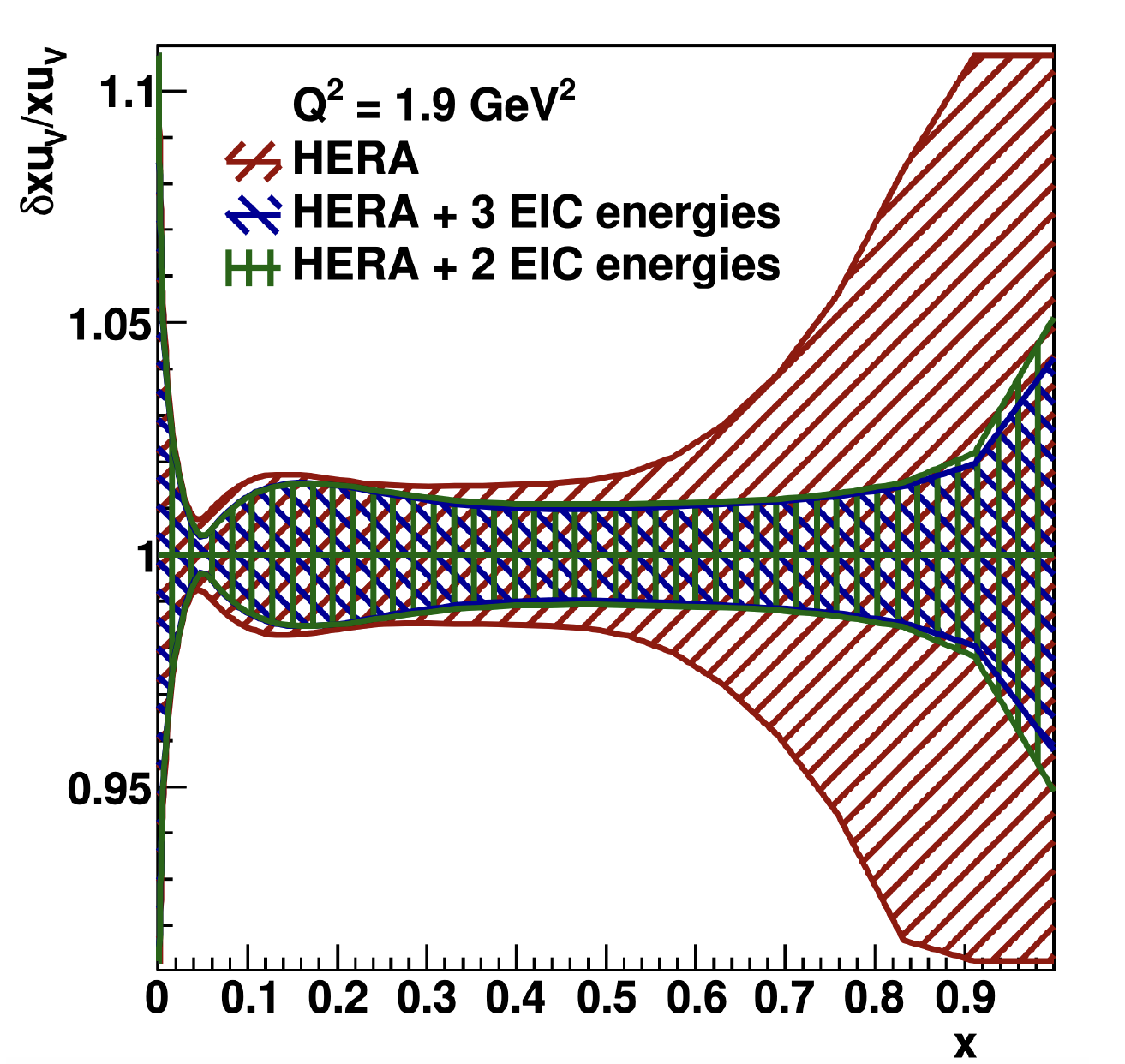}
     \includegraphics[scale = .8]{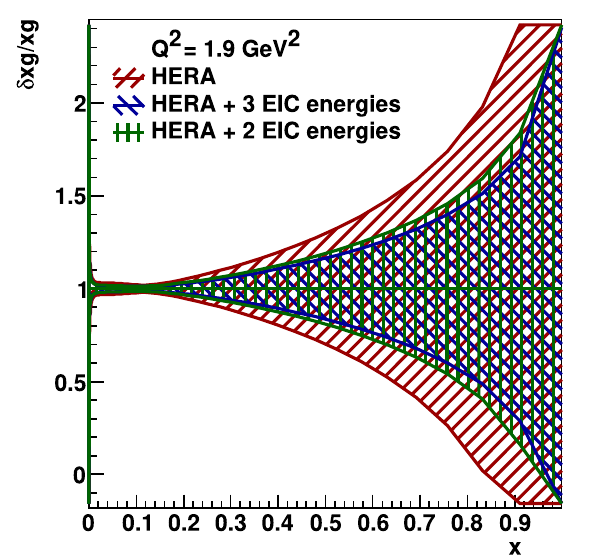}
    \caption{Impact of simulated EIC data on the NNLO collinear parton distributions of the proton (linear $x$ scale). 
     The bands show relative experimental uncertainties for the up-valence and gluon distributions 
     for $Q^2 = \SI{1.9}{\giga\electronvolt\squared}$. The HERAPDF2.0NNLO total uncertainties (using HERA data alone) are compared with results 
     in which simulated EIC data  for 
     different beam energy scenarios are also included in the HERAPDF2.0NNLO fitting framework.
     Beam settings are $\sqrt{s}=100$ and $72$ GeV for the two-beam case, with $\sqrt{s}=51$ GeV added for the three-beam case.}
    \label{pdf2}
\end{figure}

The impact of simulated early science EIC data in the two and three beam energy configurations on the NNLO collinear parton distributions of the proton is shown in Figures~\ref{pdf1} and~\ref{pdf2}, with logarithmic and linear $x$ scales, respectively, for the examples of the up-valence quarks and 
gluons.
For low-to-mid $x$ there is a modest improvement for both
parton types
with either EIC scenario.
For intermediate and high $x > 0.1$, there is a somewhat larger 
improvement for the gluons and a very substantial improvement
for
up-valence quarks with the inclusion of early science EIC data.
The down-valence quark distribution also exhibits a modest
improvement, 
whilst only very small improvements are observed for the
sea quark distributions (not shown). 
There is also very little
difference between the two and three EIC beam 
scenarios. 
When compared to to the results 
assuming full longer-term EIC capabilities, including 
five different centre-of-mass energies
and much larger integrated 
luminosities~\cite{armesto2024impact}, the early science PDFs are clearly less precise for low-to-mid $x$ and only slightly less precise for the large $x > 0.1$. 

\begin{figure}[htb]
    \centering
    \includegraphics[scale = 0.5]{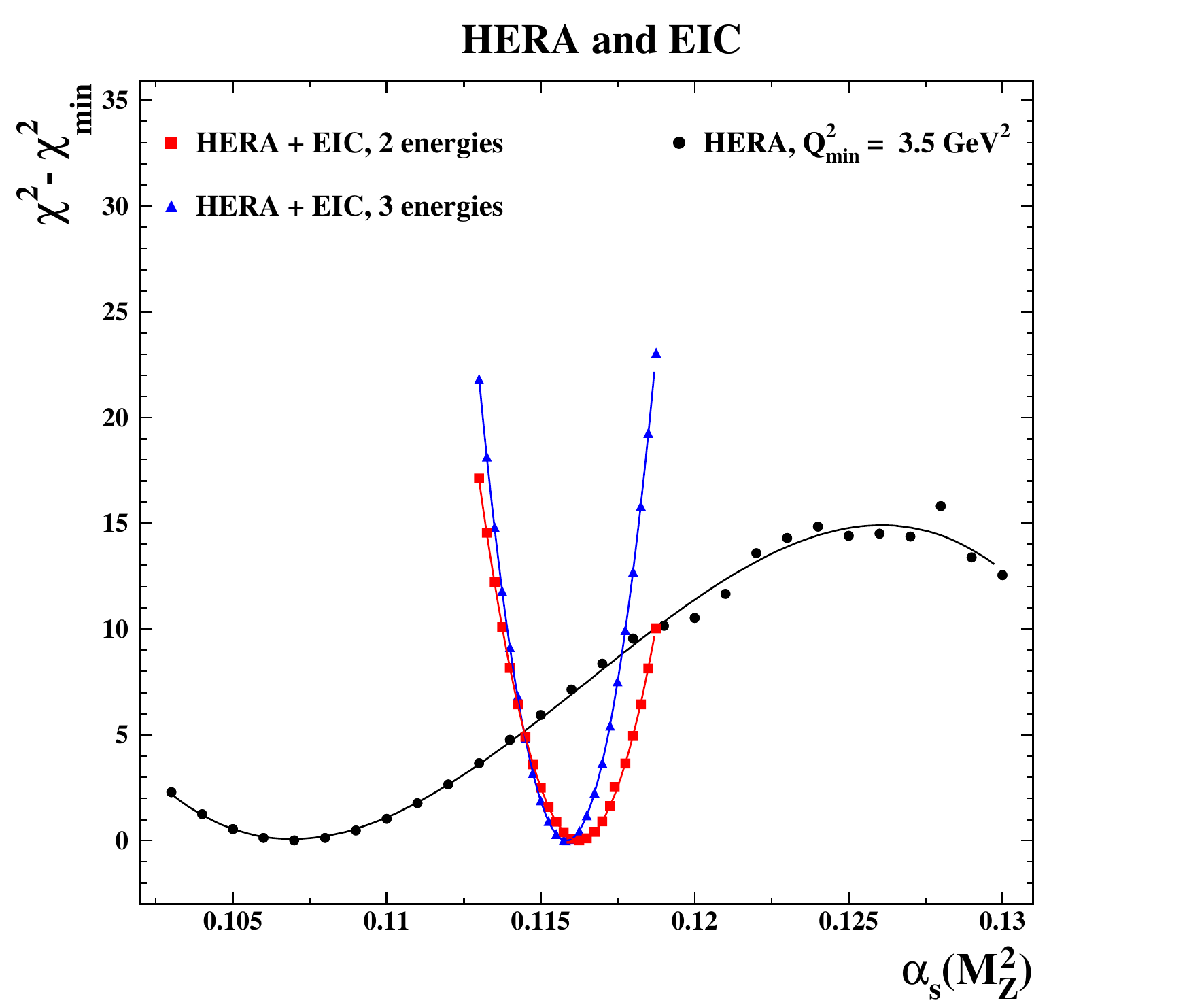}
    \caption{$\Delta\chi^2 = \chi^2 - \chi^2_{min}$ vs. $\alpha_s(M^2_{Z})$ for 
the NNLO fits to HERA data on inclusive $ep$ scattering only (black), 
and also with the addition of simulated EIC inclusive data for two (red) 
or three (blue) beam energies. 
The black full points are taken from~\cite{H1:2015ubc}.}
    \label{alphas1}
\end{figure}

A similar procedure is used to extract the strong coupling $\alpha_s$ from the inclusive HERA and EIC data, using a simultaneous fit for PDFs and $\alpha_s$, 
the only change being that
the HERAPDF2.0 variant with 
input $\alpha_s(M_Z^2) = 0.116$ is used, 
for consistency with the HERAPDF2.0Jets NNLO~\cite{H1:2021xxi}
and a previous study of the EIC sensitivity to the strong coupling~\cite{Cerci:2023uhu}. 
The procedure follows this previous publication closely.
The results and uncertainties with and without the inclusion of EIC pseudo-data are shown in
the form of a $\chi^2$ scan as a function of $\alpha_s(M_Z^2)$ 
in Fig.~\ref{alphas1}.
Each point in the figure corresponds to a full QCD fit, with all 14 PDF parameters free and a fixed
strong coupling value.
The result without EIC pseudo-data is taken from~\cite{H1:2015ubc}. 
The addition of the EIC early science data allows for a very precise extraction of the strong coupling.

The strong coupling extracted from the simultaneous fit for the PDFs and 
$\alpha_s(M^2_{Z})$, using pseudo-data for the two energies of EIC early science together with the
HERA inclusive data, is 
\begin{align}
\alpha_s(M_Z^2) = 0.1162 \pm 0.0008~(\mathrm{exp}), 
\end{align}
and that using the three energies of EIC pseudo-data is
\begin{align}
\alpha_s(M_Z^2) = 0.1158 \pm 0.0006~(\mathrm{exp}).
\end{align}

Here we quote only experimental uncertainties, since the model and parametrisation uncertainties are expected to be very small and the method for determining the theoretical uncertainties is uncertain (as discussed in~\cite{Cerci:2023uhu}). 
The $\alpha_s(M_Z^2)$ measurement is already very precise with two EIC beam configurations at the early science stage, 
reaching the level of 
the present world average or lattice QCD determinations.
Adding the third (lower) energy improves these results by \SI{\sim25}{\percent}. 
From the previous studies~\cite{Cerci:2023uhu}
the experimental uncertainty for five centre-of-mass energies 
assuming full long-term EIC capabilities in terms of luminosity and available beam energies is expected to be $\pm ~0.0004$.

\section{Alternative Scenarios}
\label{sec:alternatives}

The pseudo-data produced for the studies shown in this paper rely on assumptions for the available luminosity for each beam configuration, and also for the systematic precision that will be achieved for early science. 
Here, we briefly consider the impact of varying these assumptions on the results shown in this paper. 

The pseudo-data were produced assuming integrated luminosity of \SI{1}{\per\femto\barn} in each beam configuration. On the other hand, recent projections indicate that the annual luminosity that could be delivered in the three early science beam configurations considered here of $5\times130$, $10\times130$, and $10\times\SI{250}{\giga\electronvolt\squared}$ would be  4.36, 5.33, and \SI{9.18}{\per\femto\barn}, respectively, in the 
so-called 
high divergence configuration. Assuming equivalent systematics to those used previously, and half year science periods
with the luminosity for each configuration correspondingly 
scaled, 
no significant difference is observed in the precision of the extracted structure functions, or the uncertainties on the PDFs.
For the $\alpha_s$ extraction, the increased luminosity 
improves the precision by about 15\% and 20\% for the two and three beam scenarios, respectively. This is a result of the improved uncertainties in the large $x$ region, which is crucial for the precise extraction of the strong coupling, and is the only region where the statistical uncertainties dominate, 
thus benefitting the most from the increased luminosity.

During early science, it is likely that the 
electron-only method will be the main tool used in 
kinematic reconstruction,
as other methods that mix information from the scattered
electron and the hadronic final state
require a detailed understanding of the detector response to hadrons which may take time to achieve.
The resolution achievable using the electron method 
degrades at low $y$ as $\sim 1/ y$, which in turn leads to 
larger systematic uncertainties, for example due to corrections for 
migrations between bins. We therefore consider a scenario where the point-to-point systematic uncertainty of \SI{1.9}{\percent} degrades as $1/y$ below a threshold value of $y_\text{threshold}$, where it is assumed that this effect dominates the systematic uncertainty. The point-to-point systematic 
uncertainty $\sigma_{\text{P2P}}$ is therefore calculated as 

\begin{equation}
    \sigma_{\text{P2P}} = 
    \begin{cases}
    \frac{y_{\text{threshold}}}{y}\times 1.9\%, & y<y_{\text{threshold}} \\
    1.9\%, & y>y_{\text{threshold}}
    \end{cases}
\end{equation}

where $y_{\text{threshold}}$ is either $0.05$ or $0.1$. As before, little difference is found in the precision on the extracted structure functions, due to the lower impact of the low $y$ points in a Rosenbluth fit compared with large $y$ points. The only $F_L$ points strongly affected by this change are those at the largest $x$ for a given $Q^2$, which already have very large uncertainties.
No significant degradation in the precision of the PDFs is observed. The most impacted quantity is 
again $\alpha_s$, due to its 
sensitivty to 
the large $x$ data. With $y_\text{threshold}=0.05$, a $\sim50\%$ degradation is found, with $\delta \alpha_s=0.0009$ for the three EIC energy scenario and $\delta\alpha_s=0.0012$ for the two EIC energy case. For $y_\text{threshold}=0.1$ this increases to $\delta\alpha_s=0.0011$ and $\delta\alpha_s=0.0015$ for the three and two EIC energy scenarios, respectively.

\section{Summary}
\label{sec:conclu}

In this work, we have used simulated data to explore the potential of the Electron-Ion Collider (EIC) to measure the structure functions $F_2$ and $F_{L}$, as well as proton parton densities (PDFs) and the strong coupling  $\alpha_s(M^2_{Z})$,
in the early EIC science stage (approximately the first five years).
We have presented simulated model-independent measurements of the $F_2$ and $F_{L}$ structure functions using 
a Rosenbluth fit method, 
for two and three EIC beam configuration scenarios,
with and without the inclusion of the HERA data.
Our studies show that the EIC will be able to measure the structure functions 
$F_2$ and 
$F_{L}$ 
over a wide kinematic range in the first five years of operation,
with a meaningful precision on $F_L$ and improvements over the
already very precise HERA data for $F_2$. Our studies also show that, should a third $ep$ run with a lower centre-of-mass energy be possible during the early science period, a factor of $\sim 5$ improvement to the $F_L$ uncertainties may be achieved. 
In the PDF and $\alpha_s(M^2_{Z})$ studies, the inclusion of EIC pseudo-data for the two beam configurations 
yields a clear improvement in the PDF uncertainties, 
and allows $\alpha_s(M^2_Z)$ to be constrained with world leading precision. 
Adding the third, lower centre-of-mass energy beam configuration does not have a strong impact on the PDF extraction.
However it improves the $\alpha_s(M^2_{Z})$ early science determination by $\sim25\%$.
The studies presented here clearly show 
that EIC data will already be a highly impactful probe of perturbative Quantum Chromodynamics within
the first five years of data taking.

\section*{Acknowledgements}


\textcolor{black}{We are grateful to EIC and in
particular ePIC collaboration colleagues whose efforts towards machine, detector and software development have enabled the creation of the simulated data used here.
We also wish to thank many members of ePIC for valuable discussions and suggestions.}

\bibliography{bib}

\end{document}